\pdfoutput=1
\documentclass[aps, twocolumn, floatfix, superscriptaddress]{revtex4-1}
\usepackage{graphicx}
\DeclareGraphicsExtensions{.pdf}
\usepackage{amsmath,amssymb,bbold,bm,color}
\usepackage{float}
\usepackage{epstopdf}
\usepackage{tabularx}
\usepackage{braket}
\usepackage{booktabs}
\usepackage{array}
\usepackage{blindtext}
\usepackage{gensymb}
\usepackage{dutchcal}
\usepackage[colorlinks=true,breaklinks=true,linkcolor=blue,anchorcolor=red,citecolor=blue,urlcolor=blue]{hyperref}
\usepackage{seqsplit}
\allowdisplaybreaks[3]

\begin{document}

\title{Analytic solution to pseudo Landau levels in strongly bent graphene nanoribbons}

\author{Tianyu Liu}
\email{tliu@pks.mpg.de}
\affiliation{Max-Planck-Institut f\"ur Physik komplexer Systeme, 01187 Dresden, Germany}
\affiliation{Institute for Quantum Science and Engineering and Department of Physics, Southern University of Science and Technology, Shenzhen 518055, China}

\author{Hai-Zhou Lu}
\email{luhz@sustech.edu.cn}
\affiliation{Institute for Quantum Science and Engineering and Department of Physics, Southern University of Science and Technology, Shenzhen 518055, China}
\affiliation{Shenzhen Key Laboratory of Quantum Science and Engineering, Shenzhen 518055, China}

\begin{abstract} 
Nonuniform elastic strain is known to induce pseudo Landau levels in Dirac materials. But these pseudo Landau levels are hardly resolvable in an analytic fashion when the strain is strong, because of the emerging complicated space dependence in both the strain-modulated Fermi velocity and the strain-induced pseudomagnetic field. We here analytically characterize the solution to the pseudo Landau levels in strongly bent graphene nanoribbons, by treating the effects of the nonuniform Fermi velocity and pseudomagnetic field on equal footing. The analytic solution is detectable through the angle-resolved photoemission spectroscopy (ARPES) and allows quantitative comparison between theories and various experimental signatures of transport, such as the Shubnikov-de Haas oscillation in the complete absence of magnetic fields and the negative strain-resistivity resulting from the valley anomaly. The analytic solution can be generalized to various Dirac materials and will shed a new light on the related experimental explorations and straintronics applications. 
\end{abstract}

\date{\today}
\maketitle

\section{Introduction}
Landau levels \cite{landau1930} act as the canonical response of the orbital motion of electrons to the applied magnetic field and are the reason behind so many macroscopic quantum phenomena, such as the quantum Hall effect \cite{klitzing1980}, quantum oscillations \cite{shoenberg1984}, and quantum anomalies \cite{fukushima2008, li2016, burkov2015, son2013, huang2015, kim2013, xiong2015, zhang2016}. The formation of Landau levels in Dirac materials such as graphene or Weyl semimetals, intriguingly, does not necessarily rely on magnetic fields as long as an appropriate elastic strain is applied \cite{vozmediano2010, ilan2020, arjona2017, castro2017, roy2013, roy2014a, roy2014b, olivaleyva2020, venderbos2016, settnes2016, guinea2010a, levy2010, lu2012, li2015, yeh2011, masir2013}. Such strain displaces the Dirac cones in a space-dependent fashion analogous to magnetic fields and can thus induce low-energy pseudo Landau levels that support quantum oscillations \cite{liu2017a, liu2020a} as well as the chiral anomaly and the associated chiral magnetic effect \cite{pikulin2016, grushin2016}. In the simplest and probably the most flexible Dirac material -- graphene, the experimentally implementable strain can be as large as $27\%$ \cite{warner2012, zhang2014}, and may be of various patterns, such as bend \cite{guinea2010b, costa2012, chang2012, stuij2015}, twist \cite{zhang2014, shi2021}, and other simple uniaxial ones \cite{ho2017, lantagne2020}. 

Unfortunately, the pseudo Landau levels induced by the aforementioned strain patterns \cite{zhang2014, guinea2010b, costa2012, chang2012, stuij2015, shi2021, ho2017, lantagne2020} are \emph{dispersive} and thus are not directly interpretable by the standard Dirac theory established for the ordinary dispersionless Landau levels. For weak strain, the pseudo Landau level dispersions are often overlooked for simplicity until a recent study \cite{lantagne2020} analytically and nonperturbatively solves such dispersions in a uniaxially strained graphene nanoribbon with a nonuniform Fermi velocity but a uniform pseudomagnetic field. Nevertheless, understanding how pseudo Landau levels disperse in the presence of strong strain is a much more complicated problem remaining largely unexplored. This is presumably because the pseudo Landau levels are expected to occupy a large portion of the Brillouin zone with increased strain; and the standard procedure solving pseudo Landau levels using the linearized Hamiltonians \cite{castro2017, roy2013, venderbos2016, settnes2016, guinea2010a, guinea2010b, chang2012, ho2017, lantagne2020} at the Brillouin zone corners consequently fails. 

\begin{figure}[t]
\includegraphics[width = 8.6cm]{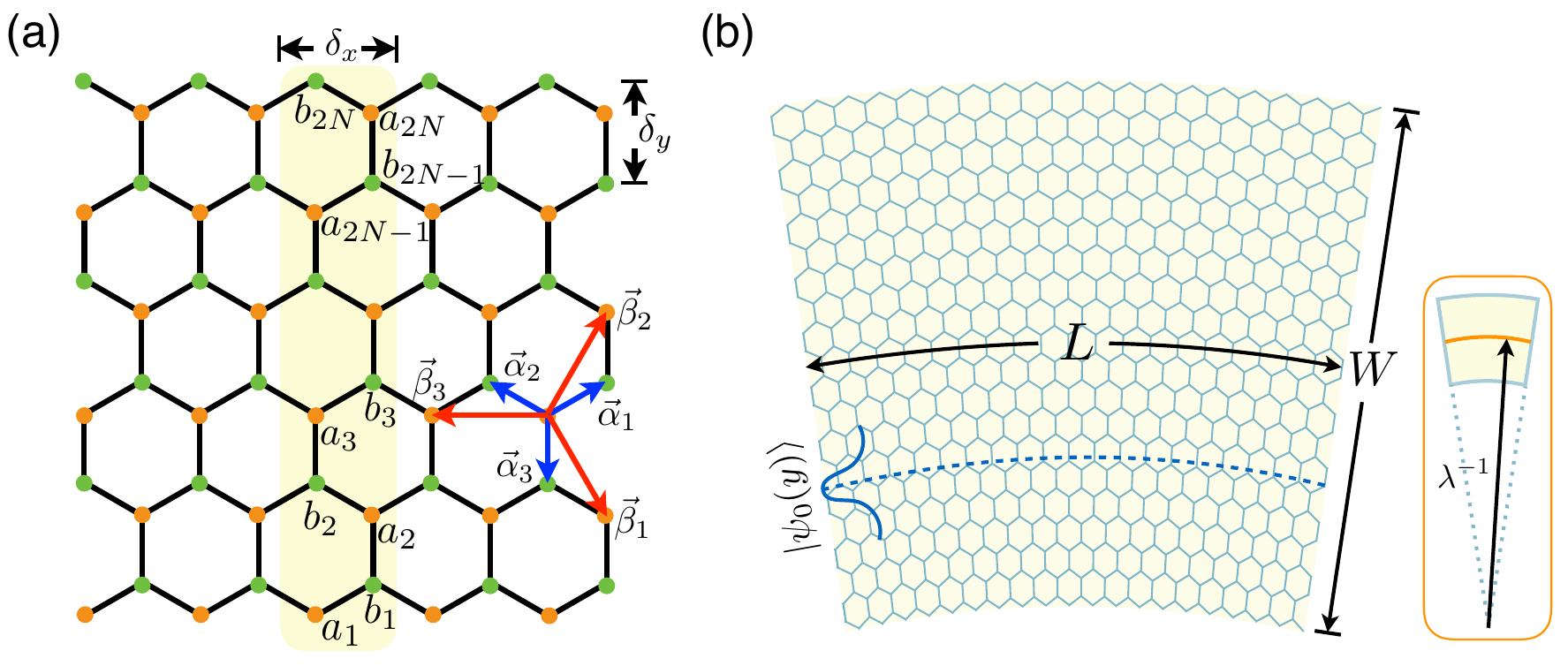}
\caption{(a) Schematic plot of an undeformed zigzag graphene nanoribbon. The yellow shadow marks the unit cell with bipartite hoppings in $a_j \leftrightarrow b_j$ and $b_j \leftrightarrow a_{j+1}$. The blue (red) arrows mark the nearest (next nearest) neighbor vectors. $\delta_{x(y)}$ is the $x(y)$ direction spacing between two neighboring sites belonging to the same sublattice. (b) Schematic plot of a fan-shaped graphene nanoribbon circularly bent from a rectangular graphene nanoribbon of length $L$ and width $W$. Note that the central arc of the bent nanoribbon coincides with the length $L$ of the undeformed nanoribbon, while the width of the bent nanoribbon is identical to its counterpart in the absence of strain. The bend may create in the bulk a domain wall (dashed) at which the bipartite hoppings are identical. The localized domain wall state is the zeroth pseudo Landau level $\ket{\psi_0(y)}$ by nature. Inset: The circular bend is parameterized by the curvature of the central arc (orange curve), denoted as $\lambda$, such that the radius of curvature of the central arc is $\lambda^{-1}$.} \label{fig1}
\end{figure}

In this paper, we present an analytic approach to solve the pseudo Landau levels in bent zigzag graphene nanoribbons under strong strain. In Sec.~\ref{sec2}, we briefly review two commonly used and analytically solvable Dirac models for weakly bent graphene nanoribbons and demonstrate the applicability as well as the limitations of such models. In Sec.~\ref{sec3}, we show that the graphene nanoribbon unit cell [Fig.~\ref{fig1}(a)] is effectively a Su-Schrieffer-Heeger model \cite{su1979} with strain-modulated bipartite hoppings, giving rise to a zero-energy topological domain wall mode [Fig.~\ref{fig1}(b)], which is actually the zeroth pseudo Landau level by nature. Linearizing the lattice model in the vicinity of the domain wall (i.e., the pseudo Landau level guiding center) into an analytically solvable Schr\"odinger differential equation, we obtain the pseudo Landau level dispersions in a wide range of the Brillouin zone. In Sec.~\ref{sec4}, we elucidate that the superiority of the lattice model over the commonly used Dirac models lies in the \emph{real-space} linearization, which treats the strain-modulated Fermi velocity and the strain-induced pseudomagnetic field on equal footing. In Sec.~\ref{sec5}, we derive the dispersions of the pseudo Landau levels for more realistic graphene models with the Semenoff mass, the intrinsic spin-orbit coupling, the electric fields, and the next nearest neighbor hoppings. The resolved analytic dispersions enable us to explore, in Sec.~\ref{sec6}, the transport resulting from the pseudo Landau levels, exemplified by the Shubnikov-de Haas oscillation in the absence of magnetic fields and the negative strain-resistivity arising from the valley anomaly. Section~\ref{sec7} concludes the paper and addresses the potential generalization of our real-space approach to a various of Dirac materials.

\section{Dirac models in the weak strain limit}
\label{sec2}
We begin by briefly reviewing the commonly used Dirac models of strained graphene \cite{castro2017, roy2013, venderbos2016, settnes2016, guinea2010a, guinea2010b, chang2012, ho2017, lantagne2020} with a focus on their applicability and limitations. In the framework of nearest neighbor tight-binding theory, the graphene Hamiltonian reads
\begin{equation} \label{tb_periodic}
H=\sum_{\bm R,i} t_i b_{\bm R + \bm \alpha_i}^\dagger a_{\bm R} + \text{H.c.},
\end{equation}
where $a_{\bm R}$ ($b_{\bm R +\bm \alpha_i}$)  annihilates an electron on the $A$ ($B$) sublattice at position $\bm R$ ($\bm R + \bm \alpha_i$) with the nearest neighbor vectors $(\bm \alpha_1, \bm \alpha_2, \bm \alpha_3) = (\tfrac{\sqrt 3}{2}a\hat x + \tfrac{1}{2}a\hat y, -\tfrac{\sqrt 3}{2}a\hat x + \tfrac{1}{2}a\hat y, -a\hat y)$ [blue arrows, Fig.~\ref{fig1}(a)] measured by the lattice constant $a=1.42\,\mathring{\text{A}}$; and $t_i$ is the electron hopping parameter between the site located at $\bm R$ and its $i$th nearest neighboring site at $\bm R+\bm \alpha_i$. In the absence of strain and anisotropy, the nearest neighbor hopping parameters are set as $t_i =t=-2.8\,\text{eV}$ \cite{castroneto2009}.

\begin{figure}[t]
\includegraphics[width = 8.6cm]{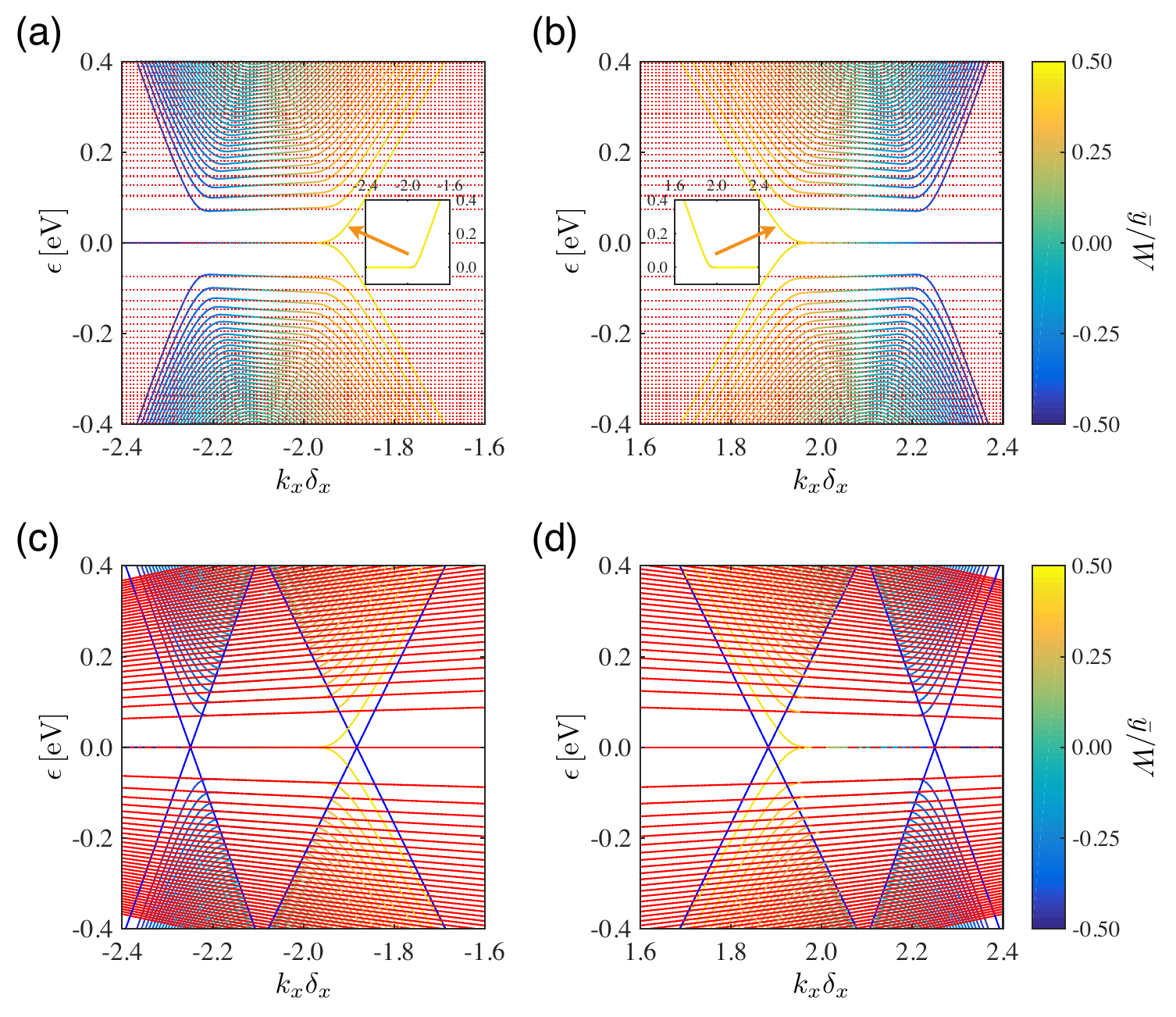}
\caption{Strain-induced pseudo Landau levels in a bent graphene nanoribbon of width $W=192\,\text{nm}$ and bend curvature $\lambda = 0.642\,\mu\text{m}^{-1}$. (a, c) and (b, d) plot the energy bands in vicinity of the left and right projected Brillouin zone corners, respectively. (a, b) Numerically calculated energy bands (solid) with theoretically predicted flat pseudo Landau levels [Eq.~(\ref{pLL_flat})] overlaid as the red dotted curves. The color scheme represents the average of the position operator $\bar y$. The insets better illustrate the marked energy bands whose flat sectors are blocked due to the degeneracy at the charge neutrality point. (c, d) The same numerical energy bands overlaid by the slightly dispersive pseudo Landau levels [Eq.~(\ref{pLL_mod_lin})] as the red solid curves. The blue curves mark the projected Dirac cones $\epsilon_{\text{max}}^{\text{DC}}$.}\label{fig2}
\end{figure}

External elastic strain alters the positions of lattice sites and thus spatially modulates the hopping parameters. In graphene, such a strain effect is incorporated through the empirical formula 
\begin{equation} \label{tsub}
t_i \rightarrow t\exp\left[ -g \tfrac{|\bm \alpha_i+\bm u(\bm R + \bm \alpha_i) - \bm u(\bm R)| - |\bm \alpha_i|}{|\bm \alpha_i|} \right], 
\end{equation}
where $\bm u(\bm r)$ is the displacement of the lattice site located at position $\bm r$ and $g=3.37$ is the Gr\"uneisen parameter \cite{pereira2009}. In the weak strain limit, the displacement field $\bm u(\bm r)$ varies slowly on the lattice scale, i.e., $|\bm \alpha_i \cdot \nabla \bm u| \ll |\bm u|$. As a common practice \cite{ilan2020, arjona2017, castro2017, settnes2016, liu2020a,  guinea2010b, ho2017, lantagne2020}, the empirical formula [Eq.~(\ref{tsub})] of the strain-modulated hopping can then be approximated by expanding to the linear order of $\nabla \bm u$ as
\begin{equation} \label{tsub_lin}
t_i \rightarrow  t \left(1- g \tfrac{ \bm \alpha_i \cdot \nabla \bm u \cdot \bm \alpha_i}{|\bm \alpha_i|^2} \right) = t(1-\tfrac{g}{a^2}\alpha_i^\mu  u_{\mu\nu}  \alpha_i^\nu),
\end{equation}
where $u_{\mu\nu}=\tfrac{1}{2}(\partial_\mu u_\nu + \partial_\nu u_\mu)$ is the strain tensor. The strain tensor should take its value at the position $\bm R+\tfrac{1}{2}\bm\alpha_i$ such that the hoppings along $\bm \alpha_i$ and $-\bm \alpha_i$ are the same. For constant strain tensors, the hopping parameters determined by Eq.~(\ref{tsub_lin}) incorporate no space dependence and the translational symmetry is preserved. By the Fourier transform $(a_{\bm r}, b_{\bm r})^T = n^{-1/2}_{\text{uc}} \sum_{\bm k} e^{i \bm k \cdot \bm r}(a_{\bm k}, b_{\bm k})^T$, where $n_{\text{uc}}$ is the number of unit cells, we obtain the Bloch Hamiltonian
\begin{equation} \label{Bloch_periodic}
\mathcal H_{\bm k} = \sum_i t_i \cos(\bm k \cdot \bm \alpha_i) \sigma^x - \sum_i t_i \sin(\bm k \cdot \bm \alpha_i) \sigma^y, 
\end{equation}
which derives from $H=\sum_{\bm k} \psi_{\bm k}^\dagger \mathcal H_{\bm k} \psi_{\bm k}$ with the sublattice basis $\psi_{\bm k} = (a_{\bm k}, b_{\bm k})^T$, where the Pauli matrices $\sigma^{x,y}$ are defined. According to Eq.~(\ref{tsub_lin}), $t_i \rightarrow t$ for weak strain. Therefore, the low-energy theory of Eq.~(\ref{Bloch_periodic}) can be obtained by linearizing $\mathcal H_{\bm k}$ in the vicinity of the Brillouin zone corners $\bm k^\eta = (\eta\tfrac{4\pi}{3 \sqrt 3a},0)$ as 
\begin{equation}
h_{\bm q}^\eta = \hbar v_x^\eta  \Big[q_x+\eta\tfrac{g (u_{yy}-u_{xx})}{2a} \Big]\sigma^x + \hbar v_y^\eta \Big(q_y+\eta\tfrac{gu_{xy}}{a} \Big)\sigma^y,
\end{equation}
where $\eta=\pm 1$ is the valley index; $(v_x^\eta, v_y^\eta) = \tfrac{3ta}{2\hbar}(-\eta,1)$ is the Fermi velocity; and $\bm q =\bm k-\bm k^\eta$ is measured from the corners of the Brillouin zone. Since $h_{\bm q}^\eta$ is in a Peierls substitution form, we can define a strain-induced vector potential $\vec {\mathcal A}^\eta = \eta \frac{g\hbar}{2ae} (u_{yy}-u_{xx}, 2u_{xy})$. Though $h_{\bm q}^\eta$ is obtained by assuming constant strain, we argue that it is in fact a legitimate theory even if the strain tensor incorporates space dependence, because the strain only varies slowly. In particular, for the weak circular bend, the displacement field reads $\bm u = \lambda (xy, -\tfrac{1}{2}x^2)$ \cite{liu2020a}, where $\lambda$ is the curvature of the central arc of the bent nanoribbon [Inset, Fig.~\ref{fig1}(b)]. Then $h_{\bm q}^\eta$ explicitly reads
\begin{equation} \label{Dirac_periodic}
h_{\bm q}^\eta = \hbar v_x^\eta (q_x - \eta \tfrac{g}{2a} \lambda  y) \sigma^x + \hbar v_y^\eta q_y \sigma^y,
\end{equation}
where a strain-induced uniform pseudomagnetic field can be defined as $\vec{\mathcal B}^\eta = \nabla \times \vec{\mathcal A}^\eta=\eta \tfrac{g\hbar}{2ea}\lambda \hat z$. The spectrum of $h_{\bm q}^\eta$  comprises of the dispersionless Dirac-Landau levels 
\begin{equation} \label{pLL_flat}
E_n^\eta = \pm \sqrt{2n\left|e\vec{\mathcal B}^\eta\hbar v_x^\eta v_y^\eta\right|} = \pm \tfrac{3}{2}t\sqrt{ng \lambda a},
\end{equation}
where the integer $n$ is the Landau level index. Equation~(\ref{pLL_flat}) is often referred to as the pseudo Landau levels in order to be distinguished from those Landau levels produced by ordinary magnetic fields. Unfortunately, Eq.~(\ref{pLL_flat}) is only capable of capturing the numerical band structure, which is obtained by diagonalizing $H$ [Eq.~(\ref{tb_periodic})] under the strain modulation $t_{1,2}=t(1-\tfrac{3}{4}g\lambda y)$ and $t_3=t$ [Eq.~(\ref{tsub_lin})], right at the projected Brillouin zone corners $k_x=\pm\mathcal k_D$ with $\mathcal k_D=\tfrac{2\pi}{3\sqrt 3a}$ [Figs.~\ref{fig2}(a) and~\ref{fig2}(b)], because $h_{\bm q}^\eta$ merely encloses the terms linear in the momentum $\bm q$ and the strain tensor $u_{xx}=\lambda y$ (note $u_{xy}=u_{yy}=0$). To improve the match, we include additional higher order terms $O(q_yq_x)$, $O(u_{xx} q_x)$, and $O(u_{xx} q_yq_x)$ to $h_{\bm q}^\eta$ and obtain a modified Dirac theory
\begin{align} \label{Dirac_mod}
\mathcal{h}_{\bm q}^\eta  =  \hbar \tilde v_x^\eta (q_x-\eta\tfrac{g}{2a}\lambda y)\sigma^x + \hbar \tilde v_y^\eta q_y \sigma^y,
\end{align}
where the renormalized Fermi velocities are $\tilde v_x^\eta = v_x^\eta (1-\tfrac{3}{4}\lambda g y)$ and $\tilde v_y^\eta = v_y^\eta(1+\tfrac{1}{2}\eta q_xa - \tfrac{1}{4}\lambda g y - \frac{3}{8}\eta q_xa \lambda g y)$, but the pseudomagnetic field is intact [cf., Eq.~(\ref{Dirac_periodic})] to the lowest order of $y$. The diagonalization of $\mathcal{h}_{\bm q}^\eta$ is analogous to the Sturm-Liouville problem analyzed in Ref.~\cite{lantagne2020} and the spectrum of $\mathcal{h}_{\bm q}^\eta$ can be analytically solved as
\begin{equation} \label{pLL_mod_lin}
\mathcal E_n^\eta(q_x) = \pm \tfrac{3}{2} t\sqrt{ng \lambda a} \sqrt{1+\tfrac{3}{2}\eta a q_x },
\end{equation}
which indeed better fits the numerical band structure in the vicinity of the projected Brillouin zone corners [Figs.~\ref{fig2}(c) and~\ref{fig2}(d)]. It is worth noting that $\mathcal E_n^\eta(q_x)$ only captures the bulk bands bounded between the two projected Dirac cones $\epsilon_{\text{max}}^{\text{DC}} = \pm \hbar \tilde v_x^\eta (q_x - \eta \tfrac{g}{2a} \lambda y)|_{y=\pm W/2}$, while the dispersive energy bands inside the projected Dirac cones and the flat energy bands emerging from the projected Dirac points are clearly originated from the marginal regions as reflected by the average of the position operator $\bar y = \int dy\, \psi_{nk_x}^*(y)\,  y\, \psi_{nk_x}(y)$, where $\psi_{nk_x}(y)$ is the wave function, as illustrated in Figs.~\ref{fig2}(c) and~\ref{fig2}(d). The real-space position of these energy bands can also be resolved by the spectral function, which is detailed in Appendix~\ref{a1}.  

We mention that $\mathcal E_n^\eta(q_x)$ derived from the modified Dirac Hamiltonian $\mathcal{h}_{\bm q}^\eta$ can gradually lose its validity when the bend curvature $\lambda$ is increased. In fact, the acquisition of $\mathcal{h}_{\bm q}^\eta$ relies on two important approximations: (i) A momentum space expansion (with respect to $\bm q$) of the Bloch Hamiltonian [Eq.~(\ref{Bloch_periodic})] in the vicinity of the Brillouin zone corners. (ii) A real space linearization (with respect to $y$) of the exponentially varying strain-modulated hopping [Eq.~(\ref{tsub})]. However, a strong strain inevitably extends the pseudo Landau levels in the momentum space and renders the momentum space expansion around the Brillouin zone corners inadequate. Moreover, the overlooked higher order terms by the linearization in the real space can become more important at strong strain. Consequently, a more sophisticated theory valid for strong strain would be desired and worthy of investigation.

\section{Lattice model in the strong strain limit}
\label{sec3}

In Sec.~\ref{sec2}, we have seen that the Dirac models are only applicable in the weak strain limit, but the expansion around the Brillouin zone corners would lose its ground for strong strain and the additional higher order terms can transform the low-energy theories to non-Dirac models, where neither the Fermi velocity nor the pseudomagnetic field can be well defined. In the present section, we develop a real-space approach based on the band topology analysis to derive the dispersions of the pseudo Landau levels induced by strong (as well as weak) circular bend.  

For the circular bend lattice deformation [Fig.~\ref{fig1}(b)], the length of the central arc coincides with the nanoribbon length before bending and the width of nanoribbon is unchanged. This implies that the azimuthal projection of a chemical bond alters linearly with the $y$ coordinate, while the radial projection of the bond is unchanged. Specifically, along the bonds $\bm \alpha_{1,2}$, the projections in the azimuthal direction become $\hat x \cdot \bm \alpha_{1,2} (1+\lambda y)$. But the bond $\bm \alpha_3$ remains intact. According to the empirical formula [Eq.~(\ref{tsub})], the modulated hopping parameters are
\begin{equation} \label{tsub_exp}
\begin{split}
t_{1,2} &\rightarrow t \exp \Big\{ g\Big[1- \sqrt{\tfrac{3}{4}(1+\lambda y)^2+\tfrac{1}{4}} \Big] \Big\} \equiv t(y), 
\\
t_3 &\rightarrow t
\end{split}
\end{equation}
which preserve the $x$ direction translational symmetry. We are thus able to perform the partial Fourier transform $(a_{\bm r},b_{\bm r})^T = N_{\text{uc}}^{-1/2} \sum_{k_x} e^{ik_xx} (a_{k_x,y}, b_{k_x,y})^T$, where $ N_{\text{uc}}$ is the number of unit cells of the bent graphene nanoribbon [Fig.~\ref{fig1}(b)], to obtain a tight-binding Hamiltonian for the circularly bent graphene nanoribbon as
\begin{equation} \label{tb_ribbon}
H=\sum_{k_x,y} b_{k_x,y+\frac{\delta_y}{6}}^\dagger [2t(y) \cos(\tfrac{1}{2}k_x\delta_x) +t \hat s_{\delta_y}] a_{k_x,y-\frac{\delta_y}{6}} + \text{H.c.},
\end{equation}
where $\delta_x = \sqrt{3} a$, $\delta_y=\tfrac{3}{2}a$, and $\hat s_{\delta_y}$ is a shift operator satisfying $\hat s_{\delta_y} a_{k_x,y} = a_{k_x,y+\delta_y}$. At a given momentum $k_x$, the nanoribbon tight-binding Hamiltonian [Eq.~(\ref{tb_ribbon})] becomes a Su-Schrieffer-Heeger model \cite{su1979} with intracell hopping $2t(y) \cos(\tfrac{1}{2}k_x\delta_x)$ and intercell hopping $t$. Due to the $y$ dependence of the hopping parameters [Eq.~(\ref{tsub_exp})], for momenta $|k_x|\leq \mathcal k_c=\tfrac{2}{\delta_x} \arccos(\tfrac{1}{2}e^{-g/2})$, a domain wall can possibly appear at
\begin{equation}\label{domain_BZ}
l_0 = \tfrac{1}{\lambda}\Big\{ \sqrt{\tfrac{4}{3} \{ 1+g^{-1} \ln [2 \cos(\tfrac{1}{2}k_x\delta_x)] \}^2-\tfrac{1}{3}}-1 \Big\},
\end{equation}
where the two hoppings are the same, while no domain wall can exist if $\mathcal k_c<|k_x|\leq \tfrac{\pi}{\delta_x}$, in which case the intercell hopping is always overwhelmed. 

\begin{figure}[t] 
\includegraphics[width = 8.6cm]{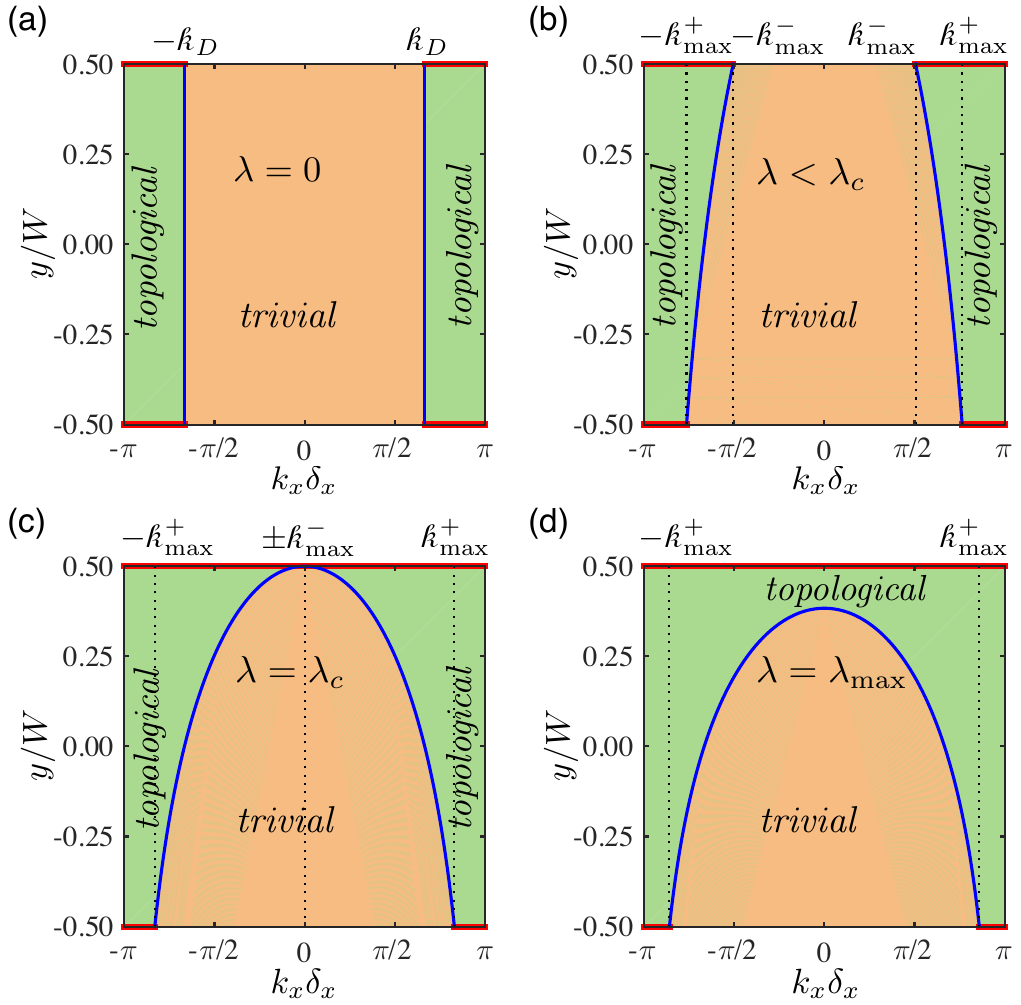}
\caption{Phase diagrams of a bent graphene nanoribbon of a generic width $W$. (a) An undeformed nanoribbon with $\lambda=0$. (b) A moderately bent nanoribbon with $\lambda W=0.263$. (c) A critically bent nanoribbon with $\lambda_c W=0.534$. (d) A maximally bent nanoribbon with $\lambda_{\text{max}} W=0.696$. In each panel, the blue curve between the dashed lines marks the position of the Su-Schrieffer-Heeger domain wall [Eq.~(\ref{domain_BZ})]; and the green (orange) patch above (below) the blue curve labels the topological (trivial) segment of the nanoribbon unit cell. The topological segments also produce edge states at the charge neutrality point as indicated by the bold red lines at both the stretched edge ($y=W/2$) and the compressed edge ($y=-W/2$). }\label{fig3}
\end{figure}

The position of the domain wall has a profound influence on the band topology of the nanoribbon tight-binding Hamiltonian [Eq.~(\ref{tb_ribbon})]. For an undeformed nanoribbon with $\lambda = 0$, the domain wall can only be located within the nanoribbon at the $k_x=\pm \mathcal k_D$. For $|k_x|>\mathcal k_D$ ($|k_x|<\mathcal k_D$), the intercell (intracell) hopping dominates and the unit cell becomes a topological (trivial) Su-Schrieffer-Heeger chain with (without) a pair of end modes. It is such end modes that constitute for the momenta $\mathcal k_D\leq |k_x| \leq \tfrac{\pi}{\delta_x}$ the well-known flat zigzag edge states [Fig.~\ref{fig3}(a)]. For a moderately bent graphene nanoribbon with $0 <\lambda <\lambda_c$, where $\lambda_c= \tfrac{2}{W}  \{[\tfrac{4}{3} (1+g^{-1} \ln 2  )^2-\tfrac{1}{3}]^{1/2}-1\} =0.534W^{-1}$, the domain wall is located within the nanoribbon at the momenta satisfying $\mathcal k_{\text{max}}^- \leq |k_x| \leq \mathcal k_{\text{max}}^+$, where $\mathcal k_{\text{max}}^\pm = \tfrac{2}{\delta_x} \arccos\{ \frac{1}{2} \exp[g(1\mp \tfrac{3}{4} \lambda W + \tfrac{3}{16}\lambda^2W^2)^{1/2}-g] \}$. For a given momentum $k_x$ is this range, the upper (lower) sector of the unit cell is topological (trivial), giving rise to an end mode and a domain wall mode at the charge neutrality point [Fig.~\ref{fig3}(b)]. The end modes at all allowed momenta, i.e., $\mathcal k_{\text{max}}^- \leq |k_x| \leq \mathcal k_{\text{max}}^+$, constitute a dispersionless energy band located at the stretched zigzag edge, while the domain wall modes result in a flat bulk band, which must be interpreted as the zeroth pseudo Landau level, since no other bulk states are expected to be dispersionless.  For the momenta $|k_x|>\mathcal k_{\text{max}}^+$ ($|k_x|<\mathcal k_{\text{max}}^-$), the unit cell realizes a purely topological (trivial) Su-Schrieffer-Heeger model [Fig.~\ref{fig3}(b)]. Therefore, a pair of flat edge states composed of Su-Schrieffer-Heeger end modes are expected at $\mathcal k_{\max}^+ < |k_x|\leq\tfrac{\pi}{\delta_x}$, which corresponds to the momentum-space scope of the edge state located at the compressed edge. As for the stretched edge, the ranges of the edge state add up to $\mathcal k_{\max}^- < |k_x|\leq\tfrac{\pi}{\delta_x}$. For a critically bent nanoribbon with $\lambda=\lambda_c$, the pseudo Landau levels from the left half and the right half of the Brillouin zone merge at the center, i.e., $\mathcal k_{\max}^-=0$; and the domain wall falls inside the nanoribbon for $|k_x|\leq \mathcal k_{\max}^+$ [Fig.~\ref{fig3}(c)]. The topological end modes on the stretched edge consequently constitute a flat band traversing the whole Brillouin zone [Fig.~\ref{fig3}(c)]. Such a flat band persists in a maximally bent nanoribbon with $\lambda$ increased to $\lambda_{\text{max}} = 0.696W^{-1}$  [Fig.~\ref{fig3}(d)], which corresponds to the maximal bond elongation $\sim 27\%$ \cite{zhang2014, warner2012}.

\begin{figure}[ht]
\includegraphics[width = 8.6cm]{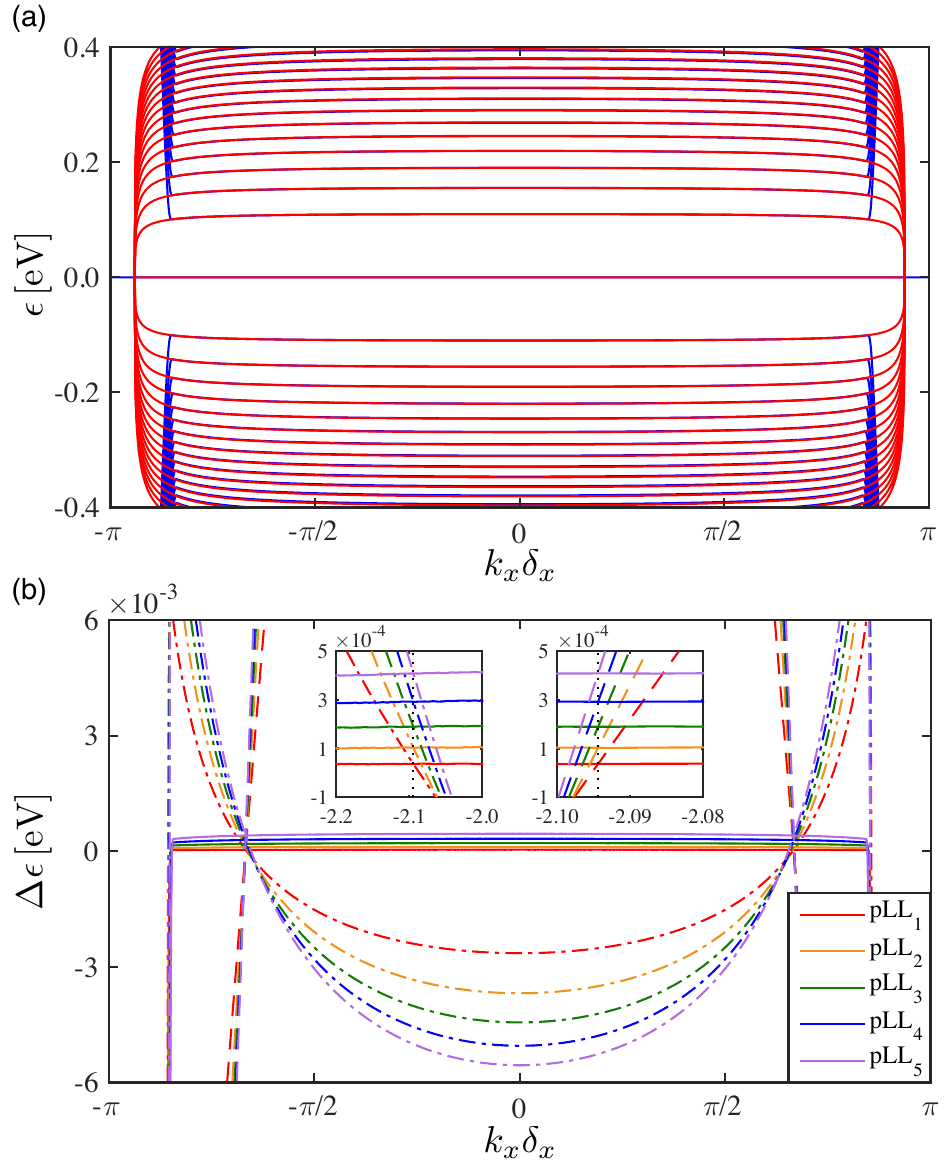}
\caption{(a) Band structure of a bent graphene nanoribbon of width $W=511\,\text{nm}$ and maximal bend curvature $\lambda_{\text{max}}=1.36\,\mu\text{m}^{-1}$. The blue curves are numerically obtained by diagonalizing the nanoribbon tight-binding Hamiltonian [Eq.~(\ref{tb_ribbon})] under the strain modulation [Eq.~(\ref{tsub_exp})]. The red curves are the dispersive pseudo Landau levels predicted by Eq.~(\ref{pLL}). (b) The energy differences between the numerical energy bands (blue) in panel (a) and various analytically proposed pseudo Landau levels [Eqs.~(\ref{pLL}),~(\ref{pLL_mod_lin}), and~(\ref{pLL_flat})] are plotted as solid, dashed, and dot dashed curves, respectively. Left (right) inset enlarges the energy differences associated with Eqs.~(\ref{pLL}) and~(\ref{pLL_flat}) [Eq~(\ref{pLL}) and~(\ref{pLL_mod_lin})] in the vicinity of $k_x=-\mathcal k_D$ (dotted line).}\label{fig4}
\end{figure}

The Su-Schrieffer-Heeger picture of the unit cell sheds new light on the resolution of the pseudo Landau levels, i.e., the spectrum of the nanoribbon Bloch Hamiltonian  
\begin{equation} \label{Bloch_ribbon}
\mathcal H_{k_x,y} = [2t(y)\cos(\tfrac{1}{2}k_x\delta_x)+t] \sigma^x -it\delta_y \sigma^y \tfrac{d}{dy}, 
\end{equation}
which is related to the nanoribbon tight-binding Hamiltonian [Eq.~(\ref{tb_ribbon})] through $H=\sum_{k_x,y} \psi_{k_x,y}^\dagger \mathcal H_{k_x,y} \psi_{k_x,y}$ with the sublattice basis $\psi_{k_x,y}=(a_{k_x,y-\delta_y/6}, b_{k_x,y+\delta_y/6})^T$. Note that we have taken the continuum limit in Eq.~(\ref{tb_ribbon}) such that the shift operator is written as $\hat s_{\delta_y} \approx 1+ \delta_y \tfrac{d}{dy}$. Because of the complicated space dependence of $t(y)$, analytically solving the Schr\"odinger differential equation characterized by $\mathcal H_{k_x,y}$ is generally not feasible. But the band topology analysis has revealed the nature of the zeroth pseudo Landau level being the Su-Schrieffer-Heeger domain wall mode, and thus locates the common guiding center of all pseudo Landau levels in the real space, provided that there are no electric fields or next nearest neighbor hoppings, whose effects are detailed in Secs.~\ref{sec5c} and~\ref{sec5d}. Since the pseudo Landau levels are well localized states, their dispersions can be in principle accurately approximated by studying the nanoribbon Bloch Hamiltonian [Eq.~(\ref{Bloch_ribbon})] in the vicinity of their common guiding center. We find it more convenient to work with the momenta $k_x \in [\tfrac{\pi}{\delta_x}, \tfrac{3\pi}{\delta_x}]$ and then maps the resolved dispersions of the pseudo Landau levels back to the conventional first Brillouin zone. Such a manipulation introduces no artifacts because the legitimate energy bands must have a $\tfrac{2\pi}{\delta_x}$ period in $k_x$, even though the nanoribbon Bloch Hamiltonian [Eq.~(\ref{Bloch_ribbon})] seemingly has a $\tfrac{4\pi}{\delta_x}$ period due to the specific form of the Fourier transform we have chosen. For the momenta $k_x \in [\tfrac{\pi}{\delta_x}, \tfrac{3\pi}{\delta_x}]$, the position of the domain wall should be rewritten as
\begin{equation}\label{domain}
\ell_0 = \tfrac{1}{\lambda} \Big\{ \sqrt{\tfrac{4}{3} \{ 1+\tfrac 1 g \ln [-2 \cos(\tfrac{1}{2}k_x\delta_x)] \}^2-\tfrac{1}{3}}-1 \Big\},
\end{equation}
which can be reduced to Eq.~(\ref{domain_BZ}) by setting $k_x \rightarrow k_x +\tfrac{2\pi}{\delta_x}$. In the vicinity of the domain wall, i.e., the common guiding center, the nanoribbon Bloch Hamiltonian [Eq.~(\ref{Bloch_ribbon})]  is restored to a standard Dirac Hamiltonian
\begin{equation} \label{Dirac_ribbon}
\mathcal h_{k_x,y} = \Omega_{\ell_0}(y-\ell_0)\sigma^x - it \delta_y  \sigma^y \tfrac{d}{dy},
\end{equation}
where $\Omega_{\ell_0} =-\tfrac{t}{t(\ell_0)} \tfrac{dt}{dy}|_{\ell_0} = \tfrac{3}{4} \lambda gt (1+\lambda \ell_0)/[\tfrac{3}{4} (1+\lambda \ell_0)^2+\tfrac{1}{4}]^{1/2}$. Alternatively, such a Dirac Hamiltonian may be written as a matrix operator
\begin{equation} \label{h0}
\mathcal h_{k_x,y} = 
\begin{bmatrix}
0 & - \epsilon_B \hat a^\dagger
\\
- \epsilon_B \hat a  & 0
\end{bmatrix},
\end{equation}
where $\epsilon_B = \sqrt{2 |\Omega_{\ell_0} t \delta_y|}$ is the energy scale. In Eq.~(\ref{h0}), $\hat a$ and $\hat a^\dagger$ are the ladder operators defined as
\begin{equation} \label{op_a}
\hat a = \tfrac{1}{\sqrt 2} (\xi_a + \tfrac{d}{d\xi_a}), \qquad \hat a^\dagger  = \tfrac{1}{\sqrt 2} (\xi_a - \tfrac{d}{d\xi_a}),
\end{equation}
in which we have defined the dimensionless parameter $\xi_a=(y-\ell_0)/l_B$ with $l_B = \sqrt{|t \delta_y/\Omega_{\ell_0}|}$ being the magnetic length. To solve the spectrum of $\mathcal h_{k_x,y}$, we adopt the trial solution $\ket {\psi_{n>0}} = (\zeta_{A,n} \ket n, \zeta_{B,n} \ket {n-1})^T$ and $\ket {\psi_0} = (\zeta_{A,0} \ket 0, 0)^T$, where $\ket n$ is defined to be an eigenstate of the bosonic number operator $\hat a^\dagger \hat a$, satisfying $\hat a^\dagger \hat a \ket n = n \ket n$. Explicitly, $\ket n$ can be written as $\ket n = (2^n \sqrt \pi n!)^{-1/2} \exp(-\xi_a^2/2) H_n(\xi_a)$, where $H_n(\cdot)$ is the $n$th Hermite polynomial. It is straightforward to verify that $\ket {\psi_{n>0}}$ ($\ket {\psi_0}$) is the eigenvector of $\mathcal h_{k_x,y}$ when $\zeta_{A,n}^2=\zeta_{B,n}^2$ ($\zeta_{A,0} \neq 0$). We here choose  $\zeta_{A,n} = \mp 1/\sqrt 2$, $\zeta_{B,n} = 1/\sqrt 2$, and $\zeta_{A,0} = 1$. And the explicit eigenvectors are
\begin{equation} \label{eigenvec}
\begin{split}
\ket{\psi_{n> 0}}=\frac{1}{\sqrt{ 2^{n+1} \pi^{\frac{1}{2}}n!}} e^{ik_xx} e^{-\frac{\xi_a^2}{2}} &\begin{bmatrix} \mp H_n(\xi_a) \\ \sqrt{2n} H_{n-1}(\xi_a) \end{bmatrix},
\\
\ket{\psi_0} = \frac{1}{\sqrt{ \pi^{\frac{1}{2}}}} e^{ik_xx} e^{-\frac{\xi_a^2}{2}} &\begin{bmatrix} H_0(\xi_a) \\ 0 \end{bmatrix},
\end{split}
\end{equation}
which correspond to the spectra $\epsilon_{n>0} = \pm \epsilon_B \sqrt{n}$ and $\epsilon_0=0$, respectively. Mapping back to the first Brillouin zone through $k_x\rightarrow k_x+\tfrac{2\pi}{\delta_x}$, we obtain the explicit dispersions of the pseudo Landau levels 
\begin{equation} \label{pLL}
\epsilon_n(k_x) =\pm \tfrac{3}{2}t\sqrt{ng\lambda a}  \sqrt[4]{\tfrac{4}{3}-\tfrac{1}{3} \tfrac{1}{\{ 1+g^{-1} \ln [2 \cos(\frac{1}{2}k_x\delta_x)] \}^2}}.
\end{equation}
Equation~(\ref{pLL}) is our key result, whose validity is justified by the good match in a wide range of momenta to the numerical band structure resulting from directly diagonalizing the nanoribbon tight-binding Hamiltonian [Eq.~(\ref{tb_ribbon})] for a maximally bent graphene nanoribbon [Fig.~\ref{fig4}(a)]. It is also worth noting that the derivation of $\epsilon_n(k_x)$ does not depend on the specific value of the bend curvature $\lambda$. Therefore, Eq.~(\ref{pLL}) is in fact applicable for both strong and weak strain. Consistent with our aforementioned analysis,  Eq.~(\ref{pLL}) is defined for $k_x\in[-\mathcal k_c, \mathcal k_c]$, in which the domain wall $l_0$ can possibly exist, while the range of the pseudo Landau levels cannot exceed the subset $[-\mathcal k_{\max}^+, \mathcal k_{\max}^+]$ in order to confine the domain wall $l_0$ inside the nanoribbon. Comparing to Eqs.~(\ref{pLL_flat}) and~(\ref{pLL_mod_lin}) derived from Dirac models [Eqs.~(\ref{Dirac_periodic}) and~(\ref{Dirac_mod})] in the weak strain limit, Eq.~(\ref{pLL}) is equally accurate at the projected Brillouin zone corners $k_x=\pm\mathcal k_D$ but exhibits much lower discrepancy with respect to the numerical band structure elsewhere for $|k_x|\leq \mathcal k_{\max}^+$ [Fig.~\ref{fig4}(b)]. 

\section{Superiority over Dirac models in the weak strain limit}
\label{sec4}
In Sec.~\ref{sec3}, we have elucidated that the dispersive pseudo Landau levels [Eq.~(\ref{pLL})] are more accurate than those [Eqs.~(\ref{pLL_flat}) and~(\ref{pLL_mod_lin})] arising from the Dirac models [Eqs.~(\ref{Dirac_periodic}) and~(\ref{Dirac_mod})] in the strong strain limit. Such a finding may not be surprising because the Dirac models are only applicable in the weak strain limit. We are thus motivated to examine whether the superiority of Eq.~(\ref{pLL}) can retain in the weak strain limit. According to Sec.~\ref{sec2}, the modified Dirac model [Eq.~(\ref{Dirac_mod})] is a more accurate low-energy theory for weak strain. We thus focus on the comparison between Eqs.~(\ref{pLL}) and~(\ref{pLL_mod_lin}) in the present section.

\begin{figure}[t]
\includegraphics[width = 8.6cm]{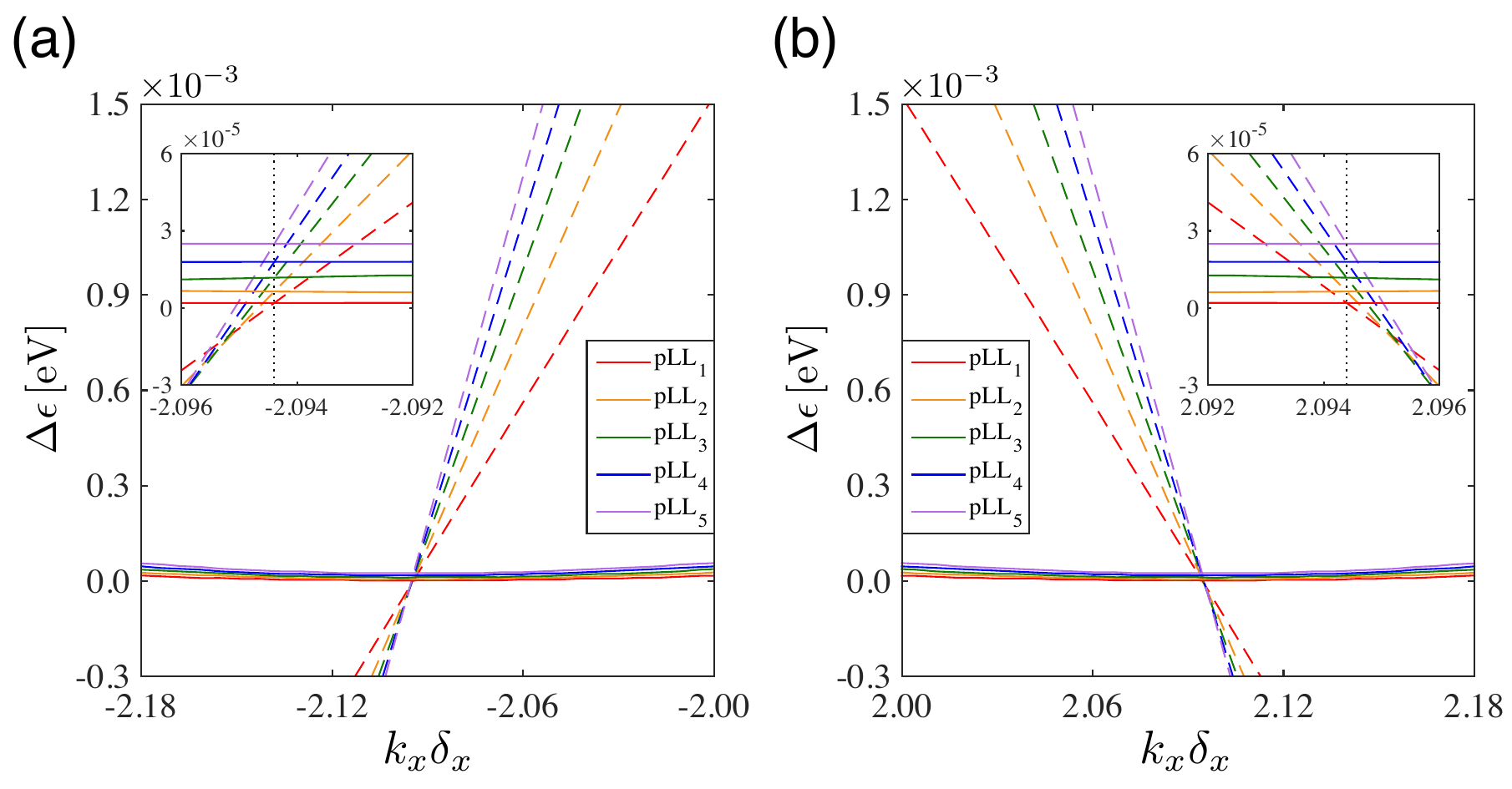}
\caption{Energy difference between the first five analytic pseudo Landau levels and the numerical energy bands for a bent graphene nanoribbon of width $W=511\,\text{nm}$ and bend curvature $\lambda=0.207\,\mu\text{m}^{-1}$. The solid (dashed) curves mark the energy difference between the pseudo Landau levels characterized by Eq.~(\ref{pLL}) [Eq.~(\ref{pLL_mod_lin})] and the numerical energy bands obtained by diagonalizing the nanoribbon tight-binding Hamiltonian [Eq.~(\ref{tb_ribbon})]. (a) Energy difference in the vicinity of $k_x=-\mathcal k_D$ with the inset enlarging the difference in a narrower range around $k_x=-\mathcal k_D$ (dotted line). (b) Energy difference in the vicinity of $k_x=\mathcal k_D$ with the inset enlarging the difference in a narrower range around $k_x=\mathcal k_D$ (dotted line). }\label{fig5}
\end{figure}

We intuitively expect Eqs.~(\ref{pLL}) and~(\ref{pLL_mod_lin}) to have similar performance in fitting the numerical band structure in the vicinity of the projected Brillouin zone corners $k_x=\pm \mathcal k_D$ for weak strain. This is because the hopping modulation [Eq.~(\ref{tsub_exp})], which is the ground for Eq.~(\ref{pLL}), can be reduced in the weak strain limit to $t_{1,2}=(1-\tfrac{3}{4}\lambda gy)$ and $t_3=t$, identical to the condition [i.e., Eq.~(\ref{tsub_lin}) with the displacement field $\bm u=\lambda(xy, -\tfrac{1} {2}x^2)$] we use to derive Eq.~(\ref{pLL_mod_lin}). Surprisingly, we find Eq.~(\ref{pLL}) exhibits much smaller deviation to the numerics than Eq.~(\ref{pLL_mod_lin}) even for weak strain [Figs.~\ref{fig5}(a) and~\ref{fig5}(b)]. Since the only difference between the two analytic dispersions $\epsilon_n(k_x)$ and $\mathcal E_n^\eta(q_x)$ lies in the hopping modulation, we thus attribute the difference to the higher order terms [e.g., $O(\lambda^2 y^2)$] overlooked during the linearization of the strain-modulated hopping $t(y)$.

To substantiate this claim, we rewrite the modified Dirac Hamiltonian [Eq.~(\ref{Dirac_mod})] as
\begin{equation}
\mathcal h_{\bm q}^\eta = \hbar v_x^\eta(y) [q_x + \tfrac{e}{\hbar} \mathcal A_x^\eta(y)] \sigma^x + \hbar v_y^\eta(y) q_y \sigma^y,
\end{equation}
with the nonuniform velocity parameters
\begin{subequations}
\begin{align}
v_x^\eta(y) &= -\tfrac{3at}{2\hbar} \eta  \tfrac{t(y)}{t},
\\
v_y^\eta(y) &= \tfrac{3at}{2\hbar}  \left[\tfrac{2}{3} + \left(\tfrac{1}{3} +\tfrac{1}{2}\eta aq_x\right) \tfrac{t(y)}{t} \right],
\end{align}
\end{subequations}
where we temporarily do not specify the space dependence of the strain-modulated hopping $t(y)$; and the strain-induced vector potential $\mathcal A^\eta_x = \tfrac{2\hbar \eta}{3ea} \tfrac{t(y)-t}{t(y)}$ gives rise to a pseudomagnetic field
\begin{equation}
\mathcal B_z^\eta(y) = - \tfrac{2\hbar \eta}{3ea} \tfrac{t \partial_y t(y)}{[t(y)]^2}.
\end{equation}
Because of the simultaneous spatial inhomogeneity in the velocity parameters and the pseudomagnetic field, the Schr\"odinger differential equation associated with $\mathcal h_{\bm q}^\eta$ is generally not analytically solvable except for $t(y)$ with simple (e.g., linear) space dependence. 

For the purpose of deriving the spectrum of $\mathcal h_{\bm q}^\eta$, we shall follow the strategy established in Sec.~\ref{sec3} by studying $\mathcal h_{\bm q}^\eta$ in the vicinity of the pseudo Landau level guiding center $y_0$, which coincides with the domain wall $\ell_0$ when $t(y)$ adopts the form of  Eq.~(\ref{tsub_exp}). For a strain-modulated hopping $t(y)$ of generic space dependence, according to Ref.~\cite{liu2020b}, the guiding center of the pseudo Landau levels is determined by $\mathcal A_x^\eta(y_0)=-\hbar q_x/e$ such that there exists a zero-energy mode in the spectrum of $\mathcal h_{\bm q}^\eta$ to be interpreted as the zeroth pseudo Landau level. By expanding in the vicinity of the guiding center $y_0$, it is straightforward to find 
\begin{equation}
\mathcal h_{\bm q}^\eta \approx -e \mathcal B_z^\eta(y_0) v_x^\eta(y_0) (y-y_0) \sigma^x + \hbar v_y^\eta(y_0) q_y \sigma^y,
\end{equation}
whose spectrum is completely determined by the velocity parameters and the pseudomagnetic field at the guiding center $y_0$. Making use of the condition $\mathcal A_x^\eta(y_0)=\tfrac{2\hbar \eta}{3ea} \tfrac{t(y_0)-t}{t(y_0)}=-\hbar q_x/e$, we find the velocity parameters are
\begin{subequations} \label{fermiv}
\begin{align}
v_x^\eta(y_0) &= -\tfrac{3at}{2\hbar} \tfrac{\eta}{1+\frac{3}{2}\eta a q_x},
\\
v_y^\eta(y_0) &=\tfrac{3at}{2\hbar},
\end{align}
\end{subequations}
which are independent of the specific space dependence of $t(y)$. However, the pseudomagnetic field sensitively depends on the form of $t(y)$ due to the appearance of $\partial_y t(y)$. Explicitly, it reads
\begin{equation} \label{pmf}
\mathcal B_z^\eta(y_0) =  \tfrac{\hbar \eta}{2ea} \lambda g \left(1+\tfrac{3}{2} \eta a q_x \right) \mathcal f_{q_x},
\end{equation}
where the coefficient reads $\mathcal f_{q_x}=1+\tfrac{3}{2} \eta a q_x$ for the linearized hopping modulation [Eq.~(\ref{tsub_lin})] and $\mathcal f_{q_x} = 1+\tfrac{1}{2g} \eta a q_x$ for the full empirical hopping modulation [Eq.~(\ref{tsub_exp})]. The resulting pseudo Landau level dispersions are
\begin{subequations}
\begin{align}
\varepsilon_n(q_x) &= \pm \tfrac{3}{2} t\sqrt{ng\lambda a} \sqrt{1+\tfrac{3}{2} \eta a q_x}, \label{pLLK1}
\\
\varepsilon_n(q_x) &= \pm \tfrac{3}{2} t\sqrt{ng\lambda a} \sqrt{1+\tfrac{1}{2g} \eta a q_x}, \label{pLLK2}
\end{align}
\end{subequations}
where the former is simply the slightly dispersive pseudo Landau levels $\mathcal E_n^\eta(q_x)$ in Eq.~(\ref{pLL_mod_lin}); and the latter corresponds to $\epsilon_n(k_x)$ in Eq.~(\ref{pLL}) expanded in the vicinity of the projected Brillouin zone corners $\eta \mathcal k_D$. Indeed, the latter is much less dispersive than the former by a ratio of $3g$, which confirms our observation in Figs.~\ref{fig5}(a) and~\ref{fig5}(b).

The finding that the higher order terms overlooked during the linearization of the strain-modulated hopping do affect the pseudomagnetic field [Eq.~(\ref{pmf})] but do not impact the Fermi velocity [Eq.~(\ref{fermiv})] up to the linear order of $q_x$ suggests that the widely used strain-modulated hoppings with linear space dependence \cite{ilan2020, arjona2017, castro2017, settnes2016, liu2020a,  guinea2010b, ho2017, lantagne2020} may be insufficient in characterizing the dispersions of the strain-induced pseudo Landau levels. To find the accurate dispersions, one would need to adopt the full space dependence of the hopping parameters without any approximation. But the complicated space dependence of such hopping parameters may hardly result in analytically solvable Schr\"odinger differential equations, which govern the dispersions of the pseudo Landau levels. In contrast, our analytic method is rooted in the band topology analysis; does not rely on the specific form of the space dependence of the hopping parameters; and thus can be transplanted to strain patterns beyond circular bend as long as such strain patterns are still characterized by $t_{1,2}\rightarrow t(y)$ and $t_3\rightarrow t$.

\section{Dispersions of pseudo Landau levels in realistic graphene}
\label{sec5}
In Sec.~\ref{sec3}, we derive the dispersions of the pseudo Landau levels using a simple nearest neighbor tight-binding model [Eq.~(\ref{tb_ribbon})] of a bent graphene nanoribbon. In realistic graphene samples, there are several inevitable effects: (i) the Semenoff mass arising from the interplay with the substrate; (ii) the Haldane mass due to the intrinsic spin-orbit coupling; (iii) the electric fields; (iv) the next nearest neighbor hoppings. The deformation of pseudo Landau levels in the presence of such effects are respectively analyzed in this section.

\subsection{Semenoff Mass}
\label{sec5a}
\begin{figure}[tb]
\includegraphics[width = 8.6cm]{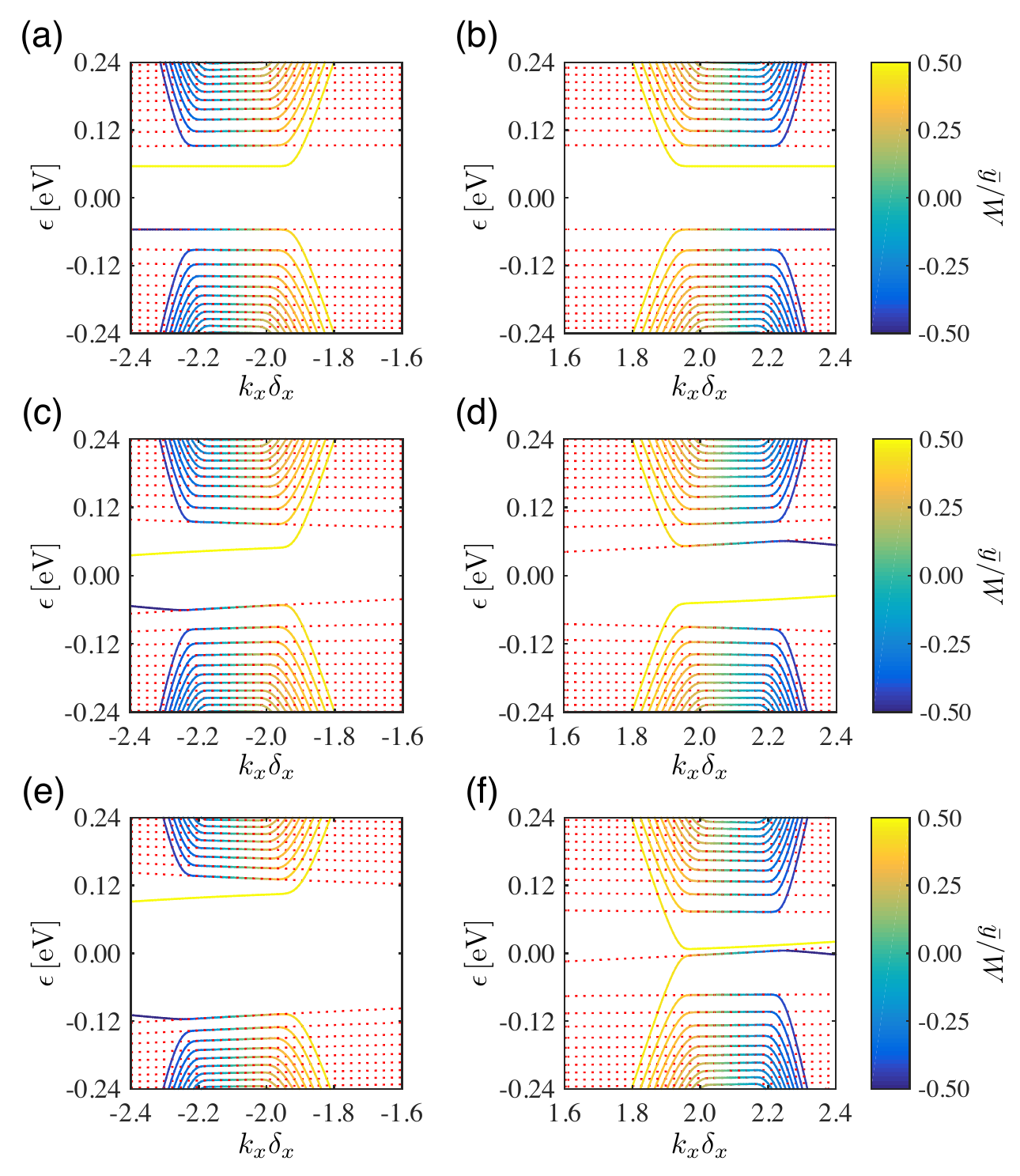}
\caption{Strain-induced pseudo Landau levels in a bent graphene nanoribbon of width $W=192\,\text{nm}$ and bend curvature $\lambda=0.642\,\mu\text{m}^{-1}$ in the presence of chiral symmetry breaking mass terms. (a, c, e) and (b, d, f) plot the energy bands in the vicinity of the left and right projected Brillouin zone corners, respectively. (a, b) The numerical band structure (solid curves) in the presence of a Semenoff mass $m_S=-56\,\text{meV}$ with analytically predicted pseudo Landau levels [Eq.~(\ref{ep_I})] overlaid as red dotted curves. (c, d) The numerical band structure (solid curves) in the presence of a Haldane mass $m_H \equiv D_{k_x, \ell_0}$ arising from the spin-orbit coupling whose strength in the absence of strain is $d'=-10.64\,\text{meV}$, such that $m_H=\pm56\,\text{meV}$ at the projected Brillouin zone corners. The analytically predicted pseudo Landau levels [Eq.~(\ref{ep_II})] are overlaid as red dotted curves. (e, f) The numerical band structure (solid curves) in the presence of both a Semenoff mass $m_S=-56\,\text{meV}$ and a Haldane mass characterized by the spin-orbit coupling $d'=-10.64\,\text{meV}$, such that the band gap at the left (right) projected Brillouin zone corner is doubled (closed). The analytically predicted pseudo Landau levels (red dotted curves) can be obtained by replacing $m_S$ in Eq.~(\ref{ep_I}) or $D_{k_x,\ell_0}$ in Eq.~(\ref{ep_II}) by an effective mass $m_{\text{eff}}=m_S+D_{k_x,\ell_0}$. For all panels,  the color index of the energy bands indicates the average value of the position operator $\bar y$.}\label{fig6}
\end{figure}

The interplay between the graphene and the substrate where it is hosted breaks the chiral symmetry by introducing a staggered potential characterized by a Semenoff mass \cite{semenoff1984}. The magnitude of the Semenoff mass $m_S$ closely relies on the details of the substrates. For hexagonal boron nitride (hBN) substrates \cite{giovannetti2007}, density functional calculations reveal $m_S=27\,\text{meV}$, while $m_S$ in silicon carbide (SiC) \cite{zhou2007, nigge2019} can be as large as $m_S=135\,\text{meV}$. Due to the presence of the Semenoff mass, the linearized Bloch Hamiltonian [Eq.~(\ref{Dirac_ribbon})] acquires an extra term and becomes
\begin{equation} \label{h_I}
\mathcal h_{k_x,y}^{\text{I}} = \mathcal h_{k_x,y} + m_S \sigma^z,
\end{equation}
which may be rewritten in terms of the ladder operators [Eq.~(\ref{op_a})]  as 
\begin{equation}
\mathcal h_{k_x,y}^{\text{I}} = 
\begin{bmatrix}
m_S & - \epsilon_B \hat a^\dagger
\\
- \epsilon_B \hat a  & -m_S
\end{bmatrix}.
\end{equation}
With the trial solution $\ket {\psi_{n>0}^{\text I}} = (\zeta_{A,n}^{\text I} \ket n, \zeta_{B,n}^{\text I} \ket {n-1})^T$ and $\ket {\psi_0^{\text I}} = (\zeta_{A,0}^{\text I} \ket 0, 0)^T$, we find that $\mathcal h_{k_x,y}^{\text{I}}$ can be diagonalized when the parameters adopt the following values $\zeta_{A,n}^{\text I} = -\text{sgn}(\epsilon_n^{\text{I}})\epsilon_B [2\epsilon_n^{\text{I}}(\epsilon_n^{\text{I}}-m_S)/n]^{-1/2}$, $\zeta_{B,n}^{\text I} = [(\epsilon_n^{\text{I}}-m_S)/2\epsilon_n^{\text{I}}]^{1/2}$, and $\zeta_{A,0}^{\text I}=1$, where the spectrum reads
\begin{equation} \label{ep_I}
\epsilon_{n>0}^{\text{I}}(k_x) = \pm \sqrt{2n |\Omega_{\ell_0} t \delta_y| + m_S^2},  \qquad \epsilon_0^{\text{I}}(k_x)=m_S.
\end{equation}
Note that the zeroth pseudo Landau level is no longer located at the charge neutrality point but is pushed to $m_S$ in the energy dimension [Figs.~\ref{fig6}(a) and~\ref{fig6}(b)]. Analysis of $\bar y$ reveals that the two segments of the zeroth pseudo Landau level are still connected by the edge state originating from the compressed edge, which has the same sublattice support, while the other edge state located on the stretched edge and originally degenerate with the zeroth pseudo Landau level in the absence of $m_S$ is now separated from the zeroth pseudo Landau level by a band gap of $2 m_S$.

\subsection{Spin-orbit coupling}
\label{sec5b}
The chiral symmetry can also be broken intrinsically by the spin-orbit coupling $H_{\text{SO}} \sim \bm s \cdot (\nabla V \times \bm k)$, where $\bm s$ is the Pauli matrix in spin space \cite{kane2005, min2006}. Such a spin-orbit coupling term further breaks the time-reversal symmetry and is known to topologically gap out the Dirac cones of graphene by introducing a Haldane mass \cite{haldane1988}, which possesses opposite signs at the different projected Brillouin zone corners. The effect of the spin-orbit coupling can be modeled by the following imaginary next nearest neighbor hopping terms
\begin{subequations} \label{tb_periodic_soc}
\begin{align}
H'_{\text{SO}_a} &= i \sum_{\bm r_a} \sum_i d'_i(\bm r_a + \tfrac{1}{2}\bm \beta_i) a_{\bm r_a + \bm \beta_i}^\dagger a_{\bm r_a} + \text{H.c.},
\\
H'_{\text{SO}_b} &= i \sum_{\bm r_b} \sum_i d'_i(\bm r_b + \tfrac{1}{2}\bm \beta_i) b_{\bm r_b}^\dagger b_{\bm r_b + \bm \beta_i} + \text{H.c.},
\end{align}
\end{subequations}
where $\bm r_a$ ($\bm r_b = \bm r_a + \bm \alpha_1$) labels the lattice sites belonging to the  $A$ ($B$) sublattice; and $(\bm \beta_1, \bm \beta_2, \bm \beta_3)=(\tfrac{\sqrt 3}{2}a\hat x - \tfrac{3}{2}a\hat y, \tfrac{\sqrt 3}{2}a\hat x + \tfrac{3}{2}a\hat y, -\sqrt 3 a \hat x)$ are the next nearest neighbor vectors [red arrows, Fig.~\ref{fig1}(a)]; and $d'_i$ measures the strength of the spin-orbit coupling associated with $\bm \beta_i$ in the presence of the circular bend. For simplicity, we assume $d_i$ to be exponentially varying, similar to the modulation of the nearest neighbor hoppings [Eq.~(\ref{tsub_exp})]. Explicitly, $d'_i$ reads
\begin{equation} \label{dpsub}
\begin{split}
d'_{1,2}(y) &= d' \exp \Big\{ g \Big[ 1-\sqrt{\tfrac{1}{4}(1+\lambda y)^2+\tfrac{3}{4}} \Big] \Big\},
\\
d'_3(y) &= d' \exp(-g\lambda y),
\end{split}
\end{equation}
where $d'$ measures the spin-orbit coupling without strain. By applying the partial Fourier transform in the $x$ direction, the nanoribbon tight-binding Hamiltonian [Eq.~(\ref{tb_ribbon})] should be supplemented by
\begin{subequations} \label{tb_ribbon_soc}
\begin{align}
H'_{\text{SO}_a} &= \sum_{k_x,y_a} a_{k_x,y_a}^\dagger D_{k_x,y_a} a_{k_x,y_a},
\\
H'_{\text{SO}_b} &= -\sum_{k_x,y_b} b_{k_x,y_b}^\dagger D_{k_x, y_b}b_{k_x,y_b},
\end{align}
\end{subequations}
where, for transparency, we have defined the parameter $D_{k_x,y} = 2\sin(\tfrac{1}{2}k_x\delta_x) [d'_1(y+\tfrac{1}{2}\delta_y) \hat s_{\delta_y} + d'_1(y-\tfrac{1}{2}\delta_y) \hat s_{-\delta_y}] - 2 d'_3(y) \sin(k_x\delta_x)$ and set $y_a=y-\tfrac{1}{6}\delta_y$ and $y_b=y+\tfrac{1}{6}\delta_y$ such that we may write Eq.~(\ref{tb_ribbon_soc}) in the sublattice basis $\psi_{k_x,y}=(a_{k_x,y-\delta_y/6}, b_{k_x,y+\delta_y/6})^T$ as $H'_{\text{SO}_a} +  H'_{\text{SO}_b} = \sum_{k_x,y} \psi_{k_x,y}^\dagger \mathcal H'_{\text{SO}} \psi_{k_x,y}$ with the correction to the nanoribbon Bloch Hamiltonian [Eq.~(\ref{Bloch_ribbon})] being a purely diagonal matrix $\mathcal H'_{\text{SO}} = \text{diag} (D_{k_x,y-\delta_y/6}, -D_{k_x,y+\delta_y/6})$. For experimentally available bend with $\lambda a \ll 1$, it is straightforward to see from Eq.~(\ref{dpsub}) that all $d'_i$ are slowly varying on the lattice scale such that $\mathcal H'_{\text{SO}}$ can be estimated through linearization as
\begin{equation} \label{Bloch_soc}
\mathcal H'_{\text{SO}} \approx D_{k_x,y} \sigma^z - \tfrac{1}{6}\delta_y \tfrac{d D_{k_x,y}}{dy} \sigma^0,
\end{equation}
where the first chiral symmetry breaking term is associated with the Haldane mass and opens up a band gap; and the second term emerges from the small separation of sublattices in the $y$ direction and shifts the energy bands in a $y$ dependent fashion. Although the parameter $D_{k_x,y}$ explicitly encloses shift operators $\hat s_{\pm \delta_y}$, it can be approximated as a purely scalar function of $y$
\begin{equation}
D_{k_x,y} \approx 4d'_1(y)\sin(\tfrac{1}{2}k_x\delta_x) - 2d'_3(y)\sin(k_x\delta_x),
\end{equation}
where we work in the continuum limit $\hat s_{\pm \delta_y} \approx 1\pm \delta_y\tfrac{d}{dy}$; take the linearization of $d_1'(y \pm \tfrac{1}{2}\delta_y)$; and only keep the lowest order terms. Since we are only interested in the low-energy pseudo Landau levels, which are localized around the domain wall $\ell_0$, it would be sufficient to study $D_{k_x,y}$ exactly at this domain wall. The resulting momentum dependent $D_{k_x,\ell_0}$ acts as the Haldane mass $m_H \equiv D_{k_x,\ell_0}$. In such an approximation, the linearized Bloch Hamiltonian [Eq.~(\ref{Dirac_ribbon})] should be rewritten as
\begin{equation} \label{h_II}
\mathcal h_{k_x,y}^{\text{II}} = \mathcal h_{k_x,y} + D_{k_x, \ell_0} \sigma^z,
\end{equation}
where we have neglected the second term in Eq.~(\ref{Bloch_soc}), because such a term only contributes at $\ell_0$ a tiny shift to the pseudo Landau levels. The spectrum of $\mathcal h_{k_x,y}^{\text{II}}$ can be directly written down by comparing to Eq.~(\ref{ep_I}) as
\begin{equation} \label{ep_II}
\epsilon_{n>0}^{\text{II}}(k_x) = \pm \sqrt{2n \Omega_{\ell_0} t \delta_y + D_{k_x,\ell_0}^2}, \qquad \epsilon_0^{\text{II}}(k_x)=D_{k_x,\ell_0},
\end{equation}
which captures the numerical simulations [Figs.~\ref{fig6}(c) and~\ref{fig6}(d)]. Note the strength of the spin-orbit coupling $d'=-10.64\,\text{meV}$ used in Figs.~\ref{fig6}(c) and~\ref{fig6}(d) is exaggerated in order to better show the band gap opened by the Haldane mass. The actual strength of the intrinsic spin-orbit coupling in graphene should be expected to be $10^{-3}\sim10^{-1}\,\text{meV}$, and thus can be in general neglected for the purpose of resolving pseudo Landau levels \cite{kane2005, min2006}. 

In contrast to the zeroth pseudo Landau level in the presence of the Semenoff mass [Figs.~\ref{fig6}(a) and~\ref{fig6}(b)], whose two segments at the left half and the right half of the Brillouin zone have identical energies consistent with the time-reversal symmetry, the zeroth pseudo Landau level under the spin-orbit coupling exhibits fundamentally different physics by emerging as a valence band at the left half of the Brillouin zone [Fig.~\ref{fig6}(c)] but as a conduction band at the right half of the Brillouin zone [Fig.~\ref{fig6}(d)]. Such positioning is dictated by the particle-hole symmetry, which is preserved because the Haldane mass is odd in both the chiral symmetry and the time-reversal symmetry. As is reflected by $\bar y$ in Figs.~\ref{fig6}(c) and~\ref{fig6}(d), the two segments of the zeroth pseudo Landau level are still connected by the edge state hosted by the compressed zigzag edge; and both edge states traverse the band gap topologically [not explicitly shown in Figs.~\ref{fig6}(c) and~\ref{fig6}(d)]. The band gap and the edge states can be manipulated by introducing an additional Semenoff mass such that the effective mass is the combination of the two types of masses as $m_{\text{eff}}=m_S+D_{k_x,\ell_0}$, which is now different around the two projected Brillouin zone corners $k_x=\pm\mathcal k_D$. The topology of the band gap depends on which type of mass is dominant. At the critical point $m_S=\pm D_{k_x,\ell_0}$, the zeroth pseudo Landau level can be pushed away from the charge neutrality point around the left projected Brillouin zone corner [Fig.~\ref{fig6}(e)] but pinned at the neutrality point around the right projected Brillouin zone corner [Fig.~\ref{fig6}(f)].

\subsection{Electric field}
\label{sec5c}
In the presence of a uniform electric field $\bm E=E \hat y$ along the $y$ direction, each of the electrons on the lattice acquires a potential energy $-e\phi(y)$, where the electric potential is chosen as $\phi(y)=-Ey-\phi_0$. The linearized Bloch Hamiltonian [Eq.~(\ref{Dirac_ribbon})] is then rewritten as
\begin{equation} \label{h_III}
\mathcal h_{k_x,y}^{\text{III}}= \mathcal h_{k_x,y} +e\phi_0\sigma^0 + eEy \sigma^0 - \tfrac{1}{6} eE\delta_y \sigma^z,
\end{equation}
which has chiral symmetry preserving onsite terms $e(\phi_0+Ey) \sigma^0$ and a mass term $- \tfrac{1}{6} eE\delta_y \sigma^z$ due to the small separation of sublattices along the  direction of the applied electric field. We write $\mathcal h_{k_x,y}^{\text{III}}$ in a matrix form as
\begin{equation} 
\mathcal h_{k_x,y}^{\text{III}} = 
\begin{bmatrix}
\epsilon_E(\hat a^\dagger + \hat a) + m & - \epsilon_B \hat a^\dagger
\\
- \epsilon_B \hat a  & \epsilon_E(\hat a^\dagger + \hat a)-m
\end{bmatrix},
\end{equation}
where we define the parameters $\epsilon_E= eEl_B/\sqrt 2$ and $m=-\tfrac{1}{6}eE\delta_y$ for transparency. To solve the eigenvalues of $\mathcal h_{k_x,y}^{\text{III}}$, we construct the following relation
\begin{equation} \label{ar_eq1}
K \ket {\psi_n^{\text {III}}} = [(\epsilon_n^{\text {III}})^2-m^2] \ket {\psi_n^{\text {III}}}, 
\end{equation}
where $\epsilon_n^{\text {III}}$ is the eigenvalue of $\mathcal h_{k_x,y}^{\text {III}}$ with respect to the eigenvector $\ket{\psi_n^{\text {III}}}$ and we have defined the auxiliary matrix operator $K = \epsilon_n^{\text {III}}(\sigma^z \mathcal h_{k_x,y}^{\text {III}} \sigma^z + \mathcal h_{k_x,y}^{\text {III}} -2m \sigma^z)- (\sigma^z \mathcal h_{k_x,y}^{\text {III}} -m)^2$ with no ladder operators in its off-diagonal entries. The dispersions of the pseudo Landau levels can then be obtained by resolving the eigenvalues of $K$. To diagonalize $K$, we apply a reversible (but not unitary) transformation to the eigenvector $\ket {\psi_n^{\text {III}}} = P \ket {\tilde \psi_n^{\text {III}}}$ with
\begin{equation}
P=\frac{1}{\sqrt{2\omega^2+2\epsilon_B \omega}}
\begin{bmatrix}
\epsilon_B + \omega & 2\epsilon_E
\\
2\epsilon_E & \epsilon_B + \omega
\end{bmatrix},
\end{equation}
where we have defined the parameter $\omega=\sqrt{ \epsilon_B^2 - 4 \epsilon_E^2}$. After the transformation, Eq.~(\ref{ar_eq1}) can be rewritten as
\begin{equation} \label{ar_eq2}
P^{-1} K P \ket {\tilde \psi_n^{\text {III}}} = [(\epsilon_n^{\text {III}})^2-m^2] \ket {\tilde \psi_n^{\text {III}}},
\end{equation}
where $P^{-1} K P$ is a purely diagonal matrix operator and reads
\begin{align} \label{K_op2}
P^{-1} K P &= \tfrac{1}{2} [(\epsilon_B^2-2\epsilon_E^2) \sigma^0 - \epsilon_B \omega \sigma^z] + 2 \epsilon_n^{\text {III}} \epsilon_E (\hat a^\dagger +\hat a) \sigma^0 \nonumber
\\
&- \epsilon_E^2 (\hat a^\dagger \hat a^\dagger + \hat a \hat a) \sigma^0 + (\epsilon_B^2-2\epsilon_E^2) \hat a^\dagger \hat a \sigma^0.
\end{align}
We now remove the terms linear in $\hat a$ and $\hat a^\dagger$ by translation
\begin{equation} \label{translation}
\begin{split}
\hat a &= \hat b - 2 \epsilon_n^{\text {III}} \epsilon_E/\omega^2,
\\
\hat a^\dagger &= \hat b^\dagger - 2 \epsilon_n^{\text {III}} \epsilon_E/\omega^2,
\end{split}
\end{equation}
where the shifted ladder operators are
\begin{equation}
\hat b = \tfrac{1}{\sqrt 2} (\xi_b + \tfrac{d}{d\xi_b}) \qquad \hat b^\dagger = \tfrac{1}{\sqrt 2} (\xi_b - \tfrac{d}{d\xi_b}),
\end{equation}
with the dimensionless parameter $\xi_b=2 \sqrt 2 \epsilon_n^{\text {III}}\epsilon_E/\omega^2 + \xi_a$. In terms of these shifted ladder operators, Eq.~(\ref{K_op2}) becomes
\begin{align} \label{K_op3}
P^{-1} \hat K P &= \tfrac{1}{2} [(\epsilon_B^2-2\epsilon_E^2) \sigma^0 - \epsilon_B \omega \sigma^z] - \tfrac{4\epsilon_E^2}{\omega^2} (\epsilon_n^{\text{III}})^2 \sigma^0 \nonumber
\\
&- \epsilon_E^2 (\hat b^\dagger \hat b^\dagger + \hat b \hat b) \sigma^0 + (\epsilon_B^2-2\epsilon_E^2) \hat b^\dagger \hat b \sigma^0.
\end{align}
We then remove the pairing ladder operators (i.e., $\hat b^\dagger \hat b^\dagger$ and $\hat b \hat b$) through the Bogoliubov transformation
\begin{equation} \label{rotation}
\begin{split}
\hat b = \hat c \sqrt{\tfrac{\epsilon_B^2-2\epsilon_E^2+\epsilon_B \omega}{2\epsilon_B \omega}} +  \hat c^\dagger \sqrt{\tfrac{\epsilon_B^2 - 2\epsilon_E^2-\epsilon_B \omega}{2\epsilon_B \omega}},
\\
\hat b^\dagger =  \hat c \sqrt{\tfrac{\epsilon_B^2 - 2\epsilon_E^2 -\epsilon_B \omega}{2\epsilon_B \omega}} + \hat c^\dagger \sqrt{\tfrac{\epsilon_B^2-2\epsilon_E^2+\epsilon_B \omega}{2\epsilon_B \omega}},
\end{split}
\end{equation}
where the rotated ladder operators are
\begin{equation}
\hat c = \tfrac{1}{\sqrt 2} (\xi_c + \tfrac{d}{d\xi_c}) \qquad \hat c^\dagger = \tfrac{1}{\sqrt 2} (\xi_c - \tfrac{d}{d\xi_c}),
\end{equation}
with the dimensionless parameter $\xi_c = \xi_b [(\epsilon_B^2 - 2\epsilon_E^2 + \epsilon_B\omega)^{1/2}-(\epsilon_B^2 - 2\epsilon_E^2 - \epsilon_B\omega)^{1/2}] (2\epsilon_B \omega)^{-1/2} $. In terms of the these rotated ladder operators, Eq.~(\ref{K_op3}) becomes
\begin{equation} \label{K_op4}
P^{-1}\hat K P = \epsilon_B \omega [\hat c^\dagger \hat c \sigma^0 + \tfrac{1}{2} (\sigma^0-\sigma^z)] - \tfrac{4\epsilon_E^2}{\omega^2} (\epsilon_n^{\text{III}})^2 \sigma^0.
\end{equation}
We plug Eq.~(\ref{K_op4}) into Eq.~(\ref{ar_eq2}) and solve the dispersions of the pseudo Landau levels to be
\begin{equation} \label{ep2}
\epsilon_{n>0}^{\text{III}} = \pm \tfrac{\omega}{\epsilon_B} \sqrt{n \omega \epsilon_B + m^2}, \qquad \epsilon_0^{\text{III}} = \tfrac{\omega}{\epsilon_B}m,
\end{equation}
where the sign of the zeroth pseudo Landau level is determined by requiring Eq.~(\ref{ep2}) to reduce to Eq.~(\ref{ep_I}) in the limit $\omega \rightarrow \epsilon_B$, or equivalently, $\epsilon_{E} \rightarrow 0$. It is worth noting that the pseudo Landau levels [Eq.~(\ref{ep2})] no longer share a common guiding center because of the shift operation in Eq.~(\ref{translation}). Nevertheless, when the electric fields are sufficiently weak with $E \ll \Omega_{\ell_0}/e$ (or, equivalently, $\epsilon_E \ll \epsilon_B$), the shift of the $n$th guiding center from the zeroth guiding center at $\ell_0$ should be much smaller than the magnetic length (i.e., $2\sqrt 2 \epsilon_n^{\text{III}}\epsilon_E l_B/\omega^2 \ll l_B$). And our theory $\mathcal h_{k_x,y}^{\text{III}}$ [Eq.~(\ref{h_III})] relying on the linearized Bloch Hamiltonian [Eq.~(\ref{Dirac_ribbon})] is still legitimate.

\begin{figure}[tb] 
\includegraphics[width = 8.6cm]{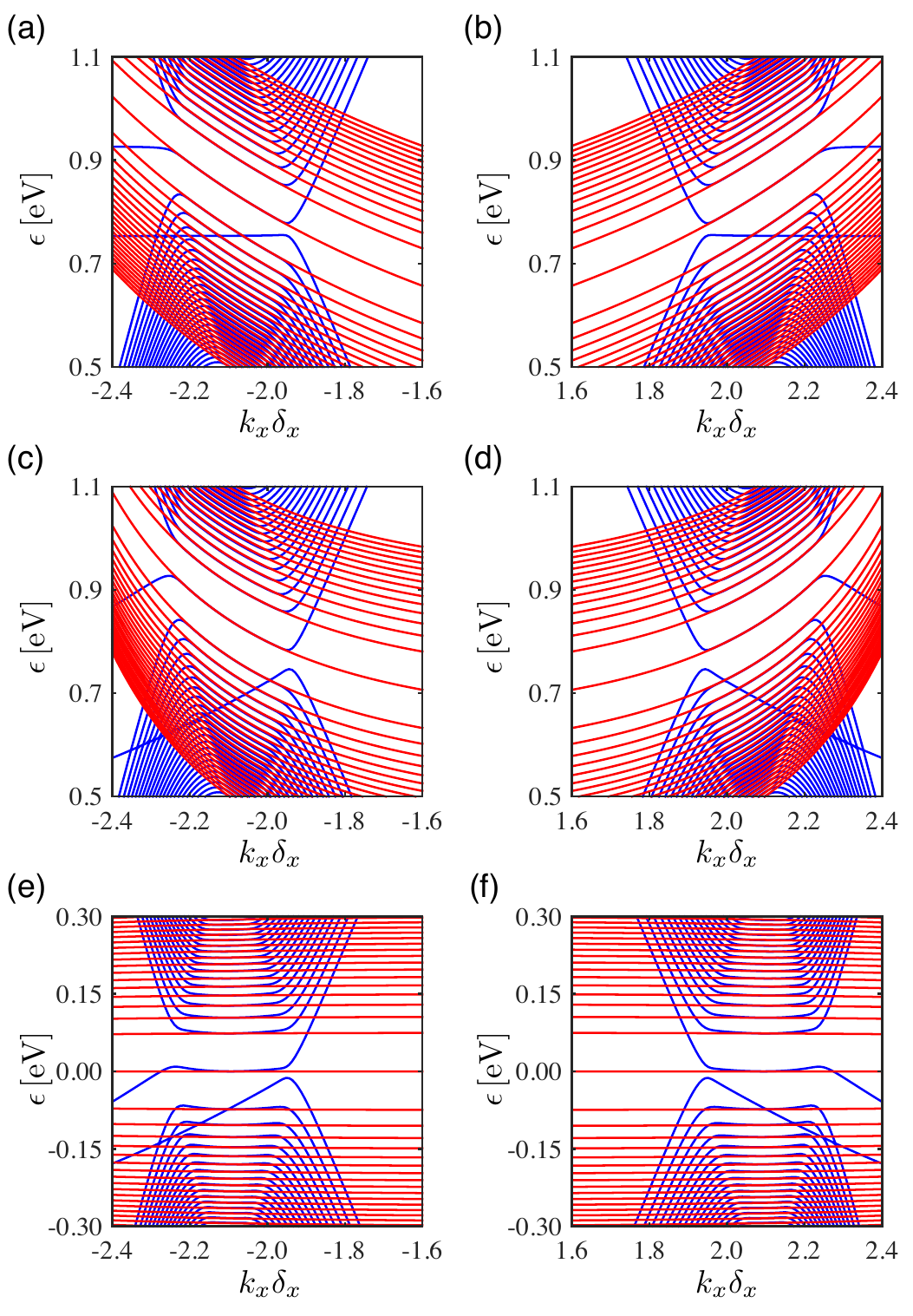}
\caption{Strain-induced pseudo Landau levels in a bent graphene nanoribbon of width $W=192\,\text{nm}$ and bend curvature $\lambda=0.642\,\mu\text{m}^{-1}$ in the presence of electric fields and/or next nearest neighbor hoppings. (a, c, e) and (b, d, f) plot the band structure in the vicinity of the left and right projected Brillouin zone corners, respectively. (a, b) The numerical band structure (blue) in the presence of a uniform electric field in the $y$ direction arising from the electric potential $\phi(y)= (0.17 \tfrac{y}{W} -0.84)\,\text{V}$. The analytically predicted pseudo Landau levels  [Eq.~(\ref{ep_III})] are overlaid as red curves. (c, d) The numerical band structure (blue) with next nearest neighbor hoppings, whose bare value in the absence of strain is $t'=-0.28\,\text{eV}$. The red curves are the predicted pseudo Landau levels [Eq.~(\ref{ep_IV})]. (e, f) The numerical band structure (blue) under both the electric potential $\phi(y)=-(0.17 \tfrac{y}{W}-0.84)\,\text{V}$ and the next nearest neighbor hoppings with $t'=-0.28\,\text{eV}$. The two effects cancel out at the projected Brillouin zone corners. And the resulting pseudo Landau levels resemble those [Eq.~(\ref{pLL})] obtained by only considering the nearest neighbor hoppings.}\label{fig7}
\end{figure}

For a weak electric field $E \ll \Omega_{\ell_0}/e$, the mass barely affects the pseudo Landau levels and can thus be safely neglected. Then Eq.~(\ref{ep2}) is reduced to
\begin{equation} \label{ep_III}
\epsilon_n^{\text{III}}(k_x)=\pm \sqrt{2n\Omega_{\ell_0}t\delta_y} \Big(1-\tfrac{e^2E^2}{\Omega_{\ell_0}^2} \Big)^{3/4} + e\phi_0+eE\ell_0.
\end{equation}
The validity of Eq.~(\ref{ep_III}) has been manifested by its accordance to the numerical simulations [Figs.~\ref{fig7}(a) and~\ref{fig7}(b)]. Such pseudo Landau levels are symmetric with respect to the Brillouin zone center because of the time-reversal symmetry, and thus are fundamentally different from the ordinary Landau levels that produce quantum Hall effects \cite{peres2007, lukose2007}.

We now briefly mention the effects of electric fields in the other two directions. A $z$ direction electric field breaks the mirror symmetry and brings up an extrinsic spin-orbit coupling Rashba term $H_R \sim \hat z \cdot (\bm s \times \bm k)$ \cite{bychkov1984}. The Rashba spin-orbit coupling arising from experimentally available electric fields is typically small comparing to the nearest neighbor hoppings \cite{huertas2006, dedkov2008, konschuh2010, zarea2009, boettger2007}, and thus should not drastically alter the strain-induced pseudo Landau levels in principle. On the other hand, an $x$ direction electric field can drive a current of electrons along the nanoribbon and lead to longitudinal transport, which will be  detailed in Sec.~\ref{sec6c}.

\subsection{Next nearest neighbor hopping}
\label{sec5d}
In realistic graphene, electrons can also hop to the next nearest neighboring sites belonging to the same sublattice and produce in the tight-binding Hamiltonian additional terms
\begin{subequations} \label{tb_periodic_nnn}
\begin{align}
H'_a &= \sum_{\bm r_a} \sum_i t'_i(\bm r_a + \tfrac{1}{2}\bm \beta_i) a_{\bm r_a + \bm \beta_i}^\dagger a_{\bm r_a} + \text{H.c.},
\\
H'_b &= \sum_{\bm r_b} \sum_i t'_i(\bm r_b + \tfrac{1}{2}\bm \beta_i) b_{\bm r_b + \bm \beta_i}^\dagger b_{\bm r_b} + \text{H.c.}.
\end{align}
\end{subequations}
Unlike the Hamiltonian [Eq.~(\ref{tb_periodic_soc})] used to model the spin-orbit coupling, the hopping parameters in Eq.~(\ref{tb_periodic_nnn}) are chosen to be purely real and exponentially varying as
\begin{equation} \label{tpsub}
\begin{split}
t'_{1,2}(y) &= t' \exp \Big\{ g \Big[ 1-\sqrt{\tfrac{1}{4}(1+\lambda y)^2+\tfrac{3}{4}} \Big] \Big\},
\\
t'_3(y) &= t' \exp(-g\lambda y),
\end{split}
\end{equation}
where $t' \in [0.02t, 0.2t]$ is the next nearest neighbor hopping in the absence of strain \cite{reich2002}. Following the procedure we have formulated in Sec.~\ref{sec5b}, it is straightforward to find out that the nanoribbon Bloch Hamiltonian [Eq.~(\ref{Bloch_ribbon})] now approximately acquires an extra term 
\begin{equation} \label{Bloch_ribbon_nnn}
\mathcal H'_{k_x,y} \approx T_{k_x,y} \sigma^0 - \tfrac{1}{6}\delta_y \tfrac{d T_{k_x,y}}{dy} \sigma^z,
\end{equation}
where we have made use of the fact that $t'_i$ are slowly varying on the lattice scale when $\lambda a \ll 1$ and defined parameter $ T_{k_x,y} = 2 \cos(\tfrac{1}{2}k_x\delta_x) [t'_1(y+\tfrac{1}{2}\delta_y)  \hat s_{\delta_y} + t'_1(y-\tfrac{1}{2}\delta_y) \hat s_{-\delta_y}] + 2t'_3(y) \cos(k_x\delta_x)$. We notice that $\mathcal H'_{k_x,y}$ resembles the terms [cf. Eq.~(\ref{h_III})] induced by an electric field $\bm E = E\hat y$ with $ T_{k_x,y}$ playing the role of the potential energy $-e\phi(y)=e\phi_0+eEy$. Although the parameter $ T_{k_x,y}$ contains shift operators $\hat s_{\pm \delta_y}$ and thus is different from the electrostatic energy, it is straightforward to show that such a parameter is approximately a purely scalar function of $y$ as
\begin{equation}
T_{k_x,y} \approx 4t'_1(y) \cos(\tfrac{1}{2}k_x\delta_x) + 2t'_3(y) \cos(k_x\delta_x),
\end{equation} 
where the continuum limit of $\hat s_{\pm \delta_y}$ and linearization of $t'_1(y\pm \tfrac{1}{2}\delta_y)$ are taken. For our purpose of finding the dispersions of low-energy pseudo Landau levels, it would be sufficient to study $T_{k_x,y}$ in the vicinity of the domain wall $\ell_0$ through the linearization $T_{k_x,y} = T_{k_x,\ell_0} + \mathcal T_{\ell_0}(y-\ell_0)$, where the derivative $\mathcal T_{\ell_0} = \tfrac{dT_{k_x,y}}{dy}|_{\ell_0} = -\lambda g \{t'_1(\ell_0) (1+\lambda \ell_0) \cos(\tfrac{1}{2}k_x\delta_x) / [\tfrac{1}{4}(1+\lambda \ell_0)^2+\tfrac{3}{4}]^{1/2} + 2t'_3(\ell_0) \cos(k_x\delta_x) \}$. Consequently, the linearized Bloch Hamiltonian [Eq.~(\ref{Dirac_ribbon})] should be rewritten as
\begin{equation} \label{h_IV}
\mathcal h_{k_x,y}^{\text{IV}} = \mathcal h_{k_x,y} + T_{k_x, \ell_0} \sigma^0 + \mathcal T_{\ell_0}(y-\ell_0) \sigma^0 - \tfrac{1}{6}\delta_y \mathcal T_{\ell_0} \sigma^z,
\end{equation}
which is analogous to Eq.~(\ref{h_III}) with $\mathcal T_{\ell_0}$ in place of the force $eE$. By comparing to Eq.~(\ref{ep_III}), we can immediately write down the pseudo Landau levels
\begin{equation} \label{ep_IV}
\epsilon_n^{\text{IV}}(k_x) = \pm \sqrt{2n \Omega_{\ell_0} t \delta_y} \Big(1-\tfrac{\mathcal T_{\ell_0}^2}{\Omega_{\ell_0}^2} \Big)^{3/4}  + T_{k_x, \ell_0},
\end{equation}
which well match the numerically calculated band structure [Figs.~\ref{fig7}(c) and~\ref{fig7}(d)]. Because of the similarity between the parameter $T_{k_x,y}$ and the potential energy $-e\phi(y)$, the next nearest neighbor effect can be exactly cancelled at (and greatly suppressed around) the projected Brillouin zone corners by an electric potential $\phi(y) = \tfrac{1}{e} [T_{k_x,\ell_0} + \mathcal T_{\ell_0} (y-\ell_0)]|_{k_x=\pm 4\pi/3\delta_x} = -\tfrac{3t'}{e}(1-\tfrac{1}{2}\lambda g y)$. Then the resulting energy bands can still be approximately characterized by Eq.~(\ref{pLL}), which is derived with only the nearest neighbor terms considered [Figs.~\ref{fig7}(e) and~\ref{fig7}(f)].

\section{Transport of bent graphene nanoribbons}
\label{sec6}
We have performed a systematic study on the analytic dispersions of pseudo Landau levels in bent graphene nanoribbons in Secs.~\ref{sec2}-\ref{sec5}. To allow comparison to experiments, analytic evaluations of transport signatures of bent graphene nanoribbons would be greatly favored. In the present section, we first justify the sufficiency of our nearest neighbor lattice model of bent graphene nanoribbons. We then phenomenologically find the analytic dispersions of the marginal energy bands spliced to the pseudo Landau levels. Ultimately, the transport signatures including the density of states (DOS), the longitudinal electrical conductivity, and the Seebeck coefficient are analytically evaluated and compared to their numerical counterparts.

\subsection{Justification of the nearest neighbor lattice model of bent graphene nanoribbons}
\label{sec6a}
In Sec.~\ref{sec5}, we have elucidated that the pseudo Landau levels resulting from the nearest neighbor lattice model [Eq.~(\ref{tb_ribbon})] are vulnerable to a variety of mechanisms such as the Semenoff mass, the spin-orbit coupling, the electric fields, and the next nearest neighbor hoppings. Despite appearing unavoidable at the first sight, these effects can actually be neglected in certain conditions. Specifically, we require a pristine graphene sample prepared on a proper substrate (e.g., hBN \cite{giovannetti2007} would be superior over SiC \cite{zhou2007, nigge2019}), where the Semenoff mass arising from the interplay with the sample is minimized; the Haldane mass (Rashba effect) resulting from the intrinsic (extrinsic) spin-orbit coupling is proved to be much smaller than the nearest neighbor hopping and can thus be neglected \cite{kane2005, min2006, huertas2006, dedkov2008, konschuh2010, zarea2009, boettger2007}; and the ubiquitous next nearest neighbor hoppings can be compensated by a properly tuned uniform $y$ direction electric field as discussed in Sec.~\ref{sec5d}. Under such conditions, it would be sufficient for the lattice model to only enclose the dominant nearest neighbor hopping terms.

Our nearest neighbor lattice model, for simplicity, only encloses the in-plane circular bend, which inhomogeneously stretches (compresses) the upper (lower) half of the nanoribbon [Fig.~\ref{fig1}(b)], while ignores the potential out-of-plane strain effects as a common practice \cite{guinea2010b, costa2012, stuij2015}. In fact, the compressive strain, even as weak as $0.1\%$ \cite{tsoukleri2009, sichen2016}, can induce out-of-plane lattice deformation (e.g., bubbles and/or wrinkles) \cite{tsoukleri2009, sichen2016, panwei2012, zhangyu2011}, which can further complicate the strain-modulated hoppings [Eq.~(\ref{tsub_exp})] by breaking the $x$ direction translational invariance. To suppress such compression-induced buckling, graphene samples should be rigidly attached to the substrate or tightly sandwiched by two substrates such that the out-of-plane lattice deformation is constrained \cite{tsoukleri2009}. To avoid the buckling in the experimental implementation, a circular bend created only by tensile strain is preferred. Such a bend can still be modeled by Eq.~(\ref{tsub_exp}) but the domain of definition of the coordinate should be adjusted to $y\in[0, W]$ from $y\in[-\tfrac{W}{2},\tfrac{W}{2}]$. Such a shift is analogous to a gauge transformation, which only relocates the guiding center but does not affect the dispersions of the pseudo Landau levels. Therefore, even for the more experimentally accessible bent graphene nanoribbons created by pure tensile strain, our key result [Eq.~(\ref{pLL})] arising from the nearest neighbor lattice model [Eq.~(\ref{tb_ribbon})] can still characterize the pseudo Landau levels.

\subsection{Phenomenological analytics of marginal energy bands}
\label{sec6b}
A full analytic analysis of the transport of bent graphene nanoribbons requires the knowledge of all energy bands. The pseudo Landau levels [Eq.~(\ref{pLL})] are clearly the bulk bands of the bent graphene nanoribbon because their common guiding center is constrained in the bulk through $-\tfrac{W}{2} \leq l_0 \leq \tfrac{W}{2}$. However, such pseudo Landau levels are distributed around $l_0$ with a characteristic width $\sim l_B$ as reflected by their wave functions [Eq.~(\ref{eigenvec})]. Therefore, when the guiding center approaches to the edges (i.e., within a few $l_B$'s), the pseudo Landau levels begin to be affected by the edges and evolve into more dispersive energy bands in the marginal regions of the nanoribbon, consistent with our observation on $\bar y$ in Figs.~\ref{fig2}(a)-\ref{fig2}(d). Such energy bands are thus referred to as the ``marginal energy bands'' in order to be distinguished from the dispersionless topological edge bands. To investigate the transport of the bent graphene nanoribbon, we aspire to quantify such marginal energy bands on a phenomenological basis. For transparency, we only consider the energy bands in the left half of the Brillouin zone, while the energy bands in the right half can be obtained by time-reversal operation.

\begin{figure*}[t] 
\includegraphics[width = 18cm]{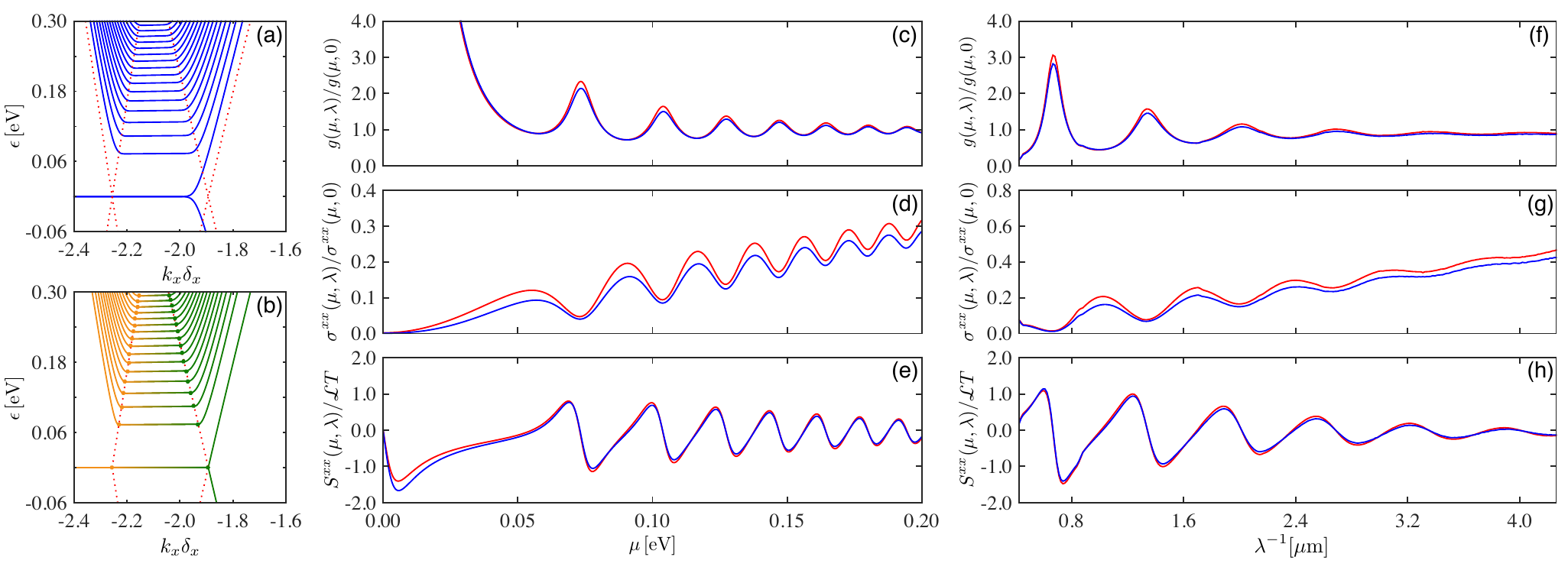}
\caption{Strain-induced quantum oscillations in a bent graphene nanoribbon of width $W=192\,\text{nm}$. (a) Numerical band structure (blue solid) at $\lambda = 0.642\,\mu\text{m}^{-1}$ in the vicinity of the left projected Brillouin zone corner at $k_x=-\mathcal k_D$ with the projected Dirac cones [Eq.~(\ref{Dirac_proj})] overlaid as dotted curves. (b) Artificial construction of the analytic band structure. The orange flat curve is the topological edge state from the compressed edge. The orange and green dispersive curves are marginal energy bands respectively characterized by Eqs.~(\ref{margin_left}) and~(\ref{margin_right}). The curves with color gradient are the pseudo Landau levels [Eq.~(\ref{pLL})]. The orange (green) scatters mark the connection of Eqs.~(\ref{pLL}) and~(\ref{margin_left}) [Eqs.(\ref{pLL}) and~(\ref{margin_right})] at the boundary of the projected Dirac cones (dotted). Quantum oscillations of (c) the DOS, (d) the electrical conductivity, (e) the Seebeck coefficient at a fixed bend curvature $\lambda = 0.642\,\mu\text{m}^{-1}$. Quantum oscillations of (f) the DOS, (g) the electrical conductivity, (h) the Seebeck coefficient at a fixed chemical potential $\mu=0.112\,\text{eV}$. In panels (c)-(h), the blue curves represent the quantities calculated from the numerical band structure in panel (a) using the tetrahedron method \cite{blochl1994}; the red curves represent the quantities calculated from the analytic band structure in panel (b) using Eqs.~(\ref{dos}),~(\ref{cond}), and~(\ref{seeb}); the units $g(\mu,0)= \tfrac{4W\mu}{9\pi a^2t^2}$ and $\sigma^{xx}(\mu, 0)=\tfrac{9e^2}{8\hbar^2}Ca^2t^2$ are respectively the DOS and the electrical conductivity in the absence of strain; the parameter $\mathcal L=2.45\times 10^{-5}\,\text{V}/\text{K}^2$ is closely related to the Lorenz number \cite{ashcroft1976}; $T$ is the temperature; and all the data are broadened by convolving in energy a Lorentzian of width $\delta_\epsilon=5.6\,\text{meV}$ to simulate the effects of disorder and finite temperature.}\label{fig8}
\end{figure*}

We first note that the width of a pseudo Landau level in the momentum space decreases with an increased Landau level index $n$ as illustrated in Fig.~\ref{fig8}(a). In fact, a higher pseudo Landau level has a more extensive wave function because of more nodes in the Hermite polynomial $H_n(\cdot)$ in Eq.~(\ref{eigenvec}), making it easier to touch the edges of the nanoribbon and thus more confined in the momentum space due to the monotonic dependence of $l_0$ on $k_x$ [cf., Figs.~\ref{fig3}(b)-\ref{fig3}(d) and Eq.~(\ref{domain_BZ})]. Consequently, all the pseudo Landau levels are phenomenologically bounded between two projected Dirac cones [Fig.~\ref{fig8}(a)]
\begin{equation} \label{Dirac_proj}
\begin{split}
\epsilon_{\text{DC}}^l (k_x) &= \pm [2t(-\tfrac{W}{2}) \cos(\tfrac{1}{2} k_x\delta_x) - t],
\\
\epsilon_{\text{DC}}^r (k_x) &=\pm [2t(\tfrac{W}{2}) \cos(\tfrac{1}{2} k_x\delta_x) - t],
\end{split}
\end{equation}
which are the projected spectra of the Bloch Hamiltonian $\mathcal H_{\bm k}$ with the parameters in Eq.~(\ref{Bloch_periodic}) set to $t_{1,2}=t(\pm\tfrac{W}{2})$, $t_3=t$, and $k_y=\pm \tfrac{2\pi}{3a}$ as well as the strong strain counterparts of $\epsilon_{\text{max}}^{\text{DC}}$. Denoting the ends of the $n$th pseudo Landau level as $k_n^{l,r}$, whose values are determined by finding the crossings of the projected Dirac cones [Eq.~(\ref{Dirac_proj})] with the pseudo Landau levels [Eq.~(\ref{pLL})], we find the real-space range of the $n$th pseudo Landau levels to be $[l_0(k_n^l), l_0(k_n^r)]$. At $l_0(k_n^{l,r})$, the pseudo Landau levels begin to evolve into marginal energy bands, whose dispersions are governed by
\begin{equation} \label{H_margin}
\tilde{\mathcal h}_{k_x,y} = \Omega_{\ell_0}(y-\ell_0)\sigma^x - it \delta_y  \sigma^y \tfrac{d}{dy}+t \left[1- \tfrac{\cos(\frac{1}{2}k_x\delta_x)}{\cos(\frac{1}{2}k_n^{l,r}\delta_x)} \right]\sigma^x,
\end{equation}
which is obtained by linearizing the nanoribbon Bloch Hamiltonian [Eq.~(\ref{Bloch_ribbon})] around $l_0(k_n^{l,r}) =\ell_0(k_n^{l,r}+\tfrac{2\pi}{\delta_x})$. The first two terms of Eq.~(\ref{H_margin}) turn out to be a Dirac Hamiltonian [cf., Eq.~(\ref{Dirac_ribbon})] characterizing pseudo Landau levels centered at $l_0(k_n^{l,r})$, while the last term can be understood as a shift to the guiding center. However, it is critically important to note that such a term must not be absorbed into the Dirac Hamiltonian, because the absorption would relocate the guiding center to somewhere outside the allowed scope of the $n$th pseudo Landau level. In the vicinity of $k_n^{l,r}$, where the linearization [Eq.~(\ref{H_margin})] of the nanoribbon Bloch Hamiltonian is legitimate, the last term in Eq.~(\ref{H_margin}) can be treated as a perturbation. Performing the perturbation calculations for $\tilde{\mathcal h}_{k_x,y}^2$, we find the first order correction to the eigenvalue to be $t^2 [1- \cos(\frac{1}{2}k_x\delta_x)/\cos(\frac{1}{2}k_n^{l,r}\delta_x)]^2$, which is doubly degenerate due to the particle-hole symmetry. Therefore, the analytic dispersions of the marginal energy bands are
\begin{subequations} \label{margin}
\begin{align}
\varepsilon_n^l(k_x) &=\sqrt {t^2\left[1-\tfrac{\cos(\frac{1}{2}k_x\delta_x)}{\cos(\frac{1}{2}k_n^l\delta_x)}\right]^2 + [\epsilon_n(k_n^l)]^2}, \label{margin_left}
\\
\varepsilon_n^r(k_x) &=\sqrt{t^2 \left[1-\tfrac{\cos(\frac{1}{2}k_x\delta_x)}{\cos(\frac{1}{2}k_n^r\delta_x)}\right]^2+[\epsilon_n(k_n^r)]^2}. \label{margin_right}
\end{align}
\end{subequations}
We note that Eqs.~(\ref{pLL}) and~(\ref{margin}) together with the flat topological edge bands at the charge neutrality point constitute an artificial band structure [Fig.~\ref{fig8}(b)] that satisfactorily mimics the numerical band structure [Fig.~\ref{fig8}(a)]. We thus expect such a band structure can phenomenologically capture the transport associated with the numerical energy bands.

\subsection{Transport signatures}
\label{sec6c}
The phenomenologically derived artificial band structure allows analytic investigation of transport signatures and comparison to numerics as well as experimental observations. Without loss of generality, we here only consider the transport of electron-like energy bands (i.e., $\mu>0$) and conduct explicit calculations in the left half of the Brillouin zone (i.e., $k_x<0$), while the transport associated with the hole-like energy bands (i.e., $\mu<0$) and the right half of the Brillouin zone (i.e., $k_x>0$) can be found using the particle-hole symmetry and the time-reversal symmetry, respectively.

We first consider the DOS of the bent graphene nanoribbon. In the bulk, the energy bands are dispersive pseudo Landau levels [Eq.~(\ref{pLL})]. The corresponding bulk DOS reads
\begin{equation} \label{dos_bulk}
g_b(\mu, \lambda)= \sum_{n \geq 0} \int_{k_n^l}^{k_n^r} \frac{dk_x}{2\pi} \delta[\epsilon_n(k_x)-\mu] = \frac{1}{2\pi}  \sum_{n \geq 0} \frac{\nu_n(\mu, \lambda)}{\frac{d\epsilon_n}{dk_x}|_{\mu}},
\end{equation}
where we define for the $n$th pseudo Landau level the occupancy parameter $\nu_n(\mu,\lambda) = \theta[\epsilon_n(k_n^r) - \mu]-\theta[\epsilon_n(k_n^r)-\mu]$ with $\theta(\cdot)$ being the Heaviside function. The dependence on the bend curvature $\lambda$ in $\nu_n(\mu,\lambda)$ is acquired from $k_n^{l,r}$. In the marginal regions, the energy bands are characterized by Eq.~(\ref{margin}), whose contribution to the DOS reads
\begin{equation} \label{dos_margin}
g_m(\mu, \lambda)=\frac{1}{2\pi} \sum_{n \geq 0} \frac{\theta[\mu-\epsilon_n(k_n^r)]}{\tfrac{d\varepsilon_n^r}{dk_x}|_\mu} - \frac{1}{2\pi} \sum_{n>0} \frac{\theta[\mu-\epsilon_n(k_n^l)]}{\tfrac{d\varepsilon_n^l}{dk_x}|_\mu}.
\end{equation}
The resulting total DOS of the bent graphene nanoribbon thus reads
\begin{equation} \label{dos}
g(\mu, \lambda)=2[g_b(\mu, \lambda)+g_m(\mu, \lambda)], 
\end{equation}
where the doubling is to include the contribution from the right half of the Brillouin zone. We find the calculated total DOS [Eq.~(\ref{dos})] satisfactorily fits the DOS numerically evaluated through the tetrahedron method \cite{blochl1994} as illustrated in Figs.~\ref{fig8}(c) and~\ref{fig8}(f) for a bent graphene nanoribbon with varying chemical potential and bend curvature, respectively. Such good matches substantiate our claim on the dispersions of the marginal energy bands [Eq.~(\ref{margin})].

We then turn to calculate the longitudinal electrical conductivity of the bent graphene nanoribbon by the Boltzmann equation approach \cite{ashcroft1976} at low temperatures (i.e., $k_BT \ll t\sqrt{g\lambda a }$). The bulk conductivity contributed by the pseudo Landau levels reads
\begin{align} \label{cond_bulk}
\sigma_b^{xx}(\mu,\lambda) &= \frac{e^2}{\hbar^2}\sum_{n\geq 0}\int_{k_n^l}^{k_n^r} \frac{dk_x}{2\pi} \tau_n^a(k_x,\lambda) \bigg(\frac{d\epsilon_n}{dk_x} \bigg)^2 \delta[\epsilon_n(k_x)-\mu] \nonumber
\\
& =\frac{e^2 \tau(\mu, \lambda)}{2\pi \hbar^2} \sum_{n \geq 0} \frac{d\epsilon_n}{dk_x}\bigg|_\mu \nu_n(\mu, \lambda),
\end{align}
where, through change of variables, we can rewrite the relaxation time  as $\tau_n^a(k_x,\lambda)=\tau_n^a(\epsilon_n^a,\lambda)$ with $\epsilon_n^a$ being the dispersion of the $n$th energy band in the artificial band structure [Fig.~\ref{fig8}(b)]. We further assume, for simplicity, an identical relaxation time $\tau_n^a(\mu, \lambda)=\tau(\mu, \lambda)$ in the second line of Eq.~(\ref{cond_bulk}). In the framework of the Fermi's golden rule, the relaxation time is inversely proportional to the DOS as $\tau(\mu, \lambda)=C/g(\mu, \lambda)$, where the proportionality coefficient $C$ encodes the information of the scattering potential in the bent graphene nanoribbon.

It is worth noting that for a certain bend curvature $\lambda$ that makes the $n$th pseudo Landau level partially occupied, i.e., $\nu_n(\mu,\lambda)=1$, the marginal energy bands always have little influence on the relaxation time because the DOS is mostly contributed by the $n$th pseudo Landau level. The bulk conductivity [Eq.~(\ref{cond_bulk})] is then reduced to $\sigma_b^{xx}(\mu,\lambda) =\tfrac{e^2C}{2\hbar^2} (\tfrac{d\epsilon_n}{dk_x})^2_\mu$, which turns out to be a decreasing function of $1/\lambda$ \cite{note1} and implies a \emph{negative} strain-resistivity analogous to the negative magnetoresistivity in the chiral magnetic effect of Weyl semimetals \cite{fukushima2008, li2016, burkov2015, son2013, huang2015, kim2013, xiong2015, zhang2016}. This negative strain-resistivity is originated from the dispersive pseudo Landau levels [Eq.~(\ref{pLL})], which play the same role as the chiral zeroth Landau levels in Weyl semimetals. Moreover, it reflects the non-conservation of the valley charge $\eta$, i.e., the valley anomaly \cite{lantagne2020}, which is a direct manifestation of the $(1+1)$-dimensional chiral anomaly \cite{nielsen1983}.

Despite the interesting anomalous transport in the bulk conductivity $\sigma_b^{xx}(\mu, \lambda)$, the major source of contribution to the total longitudinal electrical conductivity is actually from the marginal regions as
\begin{equation} \label{cond_margin}
\begin{split}
\sigma_m^{xx}(\mu, \lambda) = &- \frac{e^2 \tau(\mu, \lambda)}{2\pi \hbar^2} \sum_n \frac{d\varepsilon_n^l}{dk_x}\bigg|_\mu \theta[\mu-\epsilon_n(k_n^l)]
\\
&+\frac{e^2 \tau(\mu, \lambda)}{2\pi \hbar^2} \sum_n \frac{d\varepsilon_n^r}{dk_x}\bigg|_\mu \theta[\mu-\epsilon_n(k_n^r)],
\end{split}
\end{equation}
The total longitudinal electrical conductivity then reads
\begin{equation} \label{cond}
\sigma^{xx}(\mu, \lambda)= 2[\sigma_b^{xx}(\mu, \lambda) + \sigma_m^{xx}(\mu, \lambda)],
\end{equation}
which again encloses the contribution from the right half of the Brillouin zone. The consistency between the analytic conductivity [Eq.~(\ref{cond})] and its numerical counterpart [Figs.~\ref{fig8}(d) and~\ref{fig8}(g)] again justifies the validity of the marginal energy band dispersions [Eq.~(\ref{margin})]. For a fixed chemical potential, a scanned bend curvature can push the pseudo Landau levels through $\mu$, resulting in a periodic electron population [Fig.~\ref{fig8}(f)], which produces an unusual Shubnikov-de Haas oscillation in the complete absence of magnetic fields [Fig.~\ref{fig8}(g)]. 

With the $\mu$ dependence of $\sigma^{xx}(\mu, \lambda)$ figured out, it is straightforward to calculate the Seebeck coefficient $S^{xx}(\mu, \lambda)$ through the Mott relation \cite{cutler1969}
\begin{equation} \label{seeb}
S^{xx}(\mu,\lambda) =-\tfrac{\pi^2 k_B^2T}{3e} \tfrac{d}{d\mu}\ln \sigma^{xx}(\mu, \lambda),
\end{equation}
which is plotted in Figs.~\ref{fig8}(e) and~\ref{fig8}(h). The oscillatory behavior of the Seebeck coefficient is inherited from the longitudinal electrical conductivity [Eq.~(\ref{cond})].

\section{Conclusions}
\label{sec7}
In conclusion, the dispersions and the transport of the pseudo Landau levels in a strongly bent graphene nanoribbon are analytically studied. Such a study is motivated by the fact that the widely used Dirac models \cite{castro2017, roy2013, venderbos2016, settnes2016, guinea2010a, guinea2010b, chang2012, ho2017, lantagne2020} workable for comparatively weak strain become insufficient in the strong strain limit due to the oversimplification ignoring the nonlinear terms of the momentum and the strain tensor. Applying the band topology analysis based on the hidden chiral symmetry \cite{ryu2002}, we find that the unit cell of a bent graphene nanoribbon effectively maps to a Su-Schrieffer-Heeger model \cite{su1979} with strain-modulated bipartite hoppings. A domain wall separating the topological and trivial sectors of the unit cell results from the strain modulation and carries a zero-energy mode, which is the zeroth pseudo Landau level by nature. In the vicinity of such a domain wall (i.e., the guiding center of the pseudo Landau levels), we restore the Schr\"odinger differential equation into an analytically solvable standard Dirac equation through linearizing the model Hamiltonian. In contrast to the standard linearization adopted when deriving the Dirac models around the Brillouin zone corners, our linear expansion is conducted in real space. It thus treats the strain-modulated Fermi velocity and the strain-induced pseudomagnetic field on equal footing to give an analytic solution to the pseudo Landau levels. The resolved pseudo Landau level dispersions are accurate in a wide range of the Brillouin zone for strong strain and are even superior over the Dirac models for weak strain. 

Having acquired the dispersions of pseudo Landau levels using a nearest neighbor lattice model of bent graphene nanoribbons, we turn to consider more realistic models with chiral symmetry breaking masses, applied electric fields, and next-nearest-neighbor hoppings. The Semenoff (Haldane) mass arises from the interplay with the substrate (the intrinsic spin-orbit coupling) and opens up a trivial (topological) band gap to pseudo Landau levels. Nevertheless, comparing to the nearest neighbor hopping effect, the effect of the mass terms is comparatively small in graphene. On the other hand, the electric fields and the next nearest neighbor hoppings can be strong perturbations and thus drastically affect the electronic structure by suppressing and tilting the pseudo Landau levels. Fortunately, these two effects can cancel each other and the resulting bulk bands show no obvious difference from the pseudo Landau levels derived from the nearest neighbor lattice model. The analytically derived pseudo Landau levels and the phenomenologically approximated marginal energy bands constitute an artificial band structure allowing the analytic computation of the transport signatures (e.g., Shubnikov-de Haas oscillation in the absence of magnetic fields and the negative strain-resistivity resulting from the valley anomaly) and the comparison to numerics and experimental observations. 

Our findings may pave the way to graphene straintronics devices in the strong strain paradigm, which so far remains largely unexplored. Our approach may be transplanted to a various novel materials such as the twisted bilayer graphene \cite{liujianpeng2019}, Dirac superconductors \cite{liu2017b, kobayashi2018, matsushita2018, massarelli2017, nica2018}, and bosonic ``semimetals" \cite{liu2020b, liu2019, ferreiros2018, sunjunsong2021, sunjunsong2022, rechtsman2013, wen2019, brendel2017}, where pseudo Landau levels have been reported.

\begin{acknowledgments}
The authors are indebted to H. Scherrer-Paulus, M. Franz, P. A. McClarty, X. -X. Zhang, \'E. Lantagne-Hurtubise, T. Matsushita, S. Fujimoto, Y. Chen, H.-M. Guo, F. Peeters, and Z. Shi for insightful discussions. We particularly thank R. Moessner for the inspiring suggestions. T.L. gratefully acknowledges the scholarship from Max-Planck-Gesellschaft. H.-Z.L. is supported by the National Natural Science Foundation of China (Grants No.~\seqsplit{11534001}, No.~\seqsplit{11974249}, and No.~\seqsplit{11925402}), the National Basic Research Program of China (Grant No.~\seqsplit{2015CB921102}), Guangdong Province (Grants No.~\seqsplit{2016ZT06D348} and No.~\seqsplit{2020KCXTD001}), the National Key R \& D Program (Grant No.~\seqsplit{2016YFA0301700}), Shenzhen High-Level Special Fund (Grants No.~\seqsplit{G02206304} and No.~\seqsplit{G02206404}), and the Science, Technology and Innovation Commission of Shenzhen Municipality (Grants No.~\seqsplit{ZDSYS20170303165926217}, No.~\seqsplit{JCYJ20170412152620376}, and No.~\seqsplit{KYTDPT20181011104202253}).
\end{acknowledgments}

\appendix
\section{Spectral functions in the weak strain limit}
\label{a1}
In Sec.~\ref{sec2} of the main text, we plot the spectrum [Figs.~\ref{fig2}(a)-\ref{fig2}(d)] of a weakly bent graphene nanoribbon and use the average value of the position operator [i.e., $\bar y = \int dy\, \psi_{nk_x}^*(y)\,  y\, \psi_{nk_x}(y)$, where $\psi_{nk_x}(y)$ is the wave function] to mark the positions of the energy bands. We here show that such positions can also be resolved by the spectral function.

The spectral function of a bent graphene nanoribbon can be written as
\begin{equation} \label{spec_fn}
A_n(\epsilon, k_x) = -\frac{1}{\pi} \lim_{\delta \rightarrow 0}\Im [\epsilon + i\delta - \mathcal H_{nm}(k_x)]^{-1}_{n=m},
\end{equation}
which represents the local density of states (LDOS) at the $n$th site with $\mathcal H_{nm}(k_x)$ being the Hamiltonian matrix of the bent  graphene nanoribbon. Equation~(\ref{spec_fn}) allows us to study the spectral density in any part of the nanoribbon. For example, we may define the spectral function of the stretched (compressed) marginal region [the uppermost (lowermost) $5\%$ of the bent graphene nanoribbon] by summing $A_n(\epsilon, k_x)$ over the sites belonging to that part of the nanoribbon.

\begin{figure}[t]
\includegraphics[width = 8.6cm]{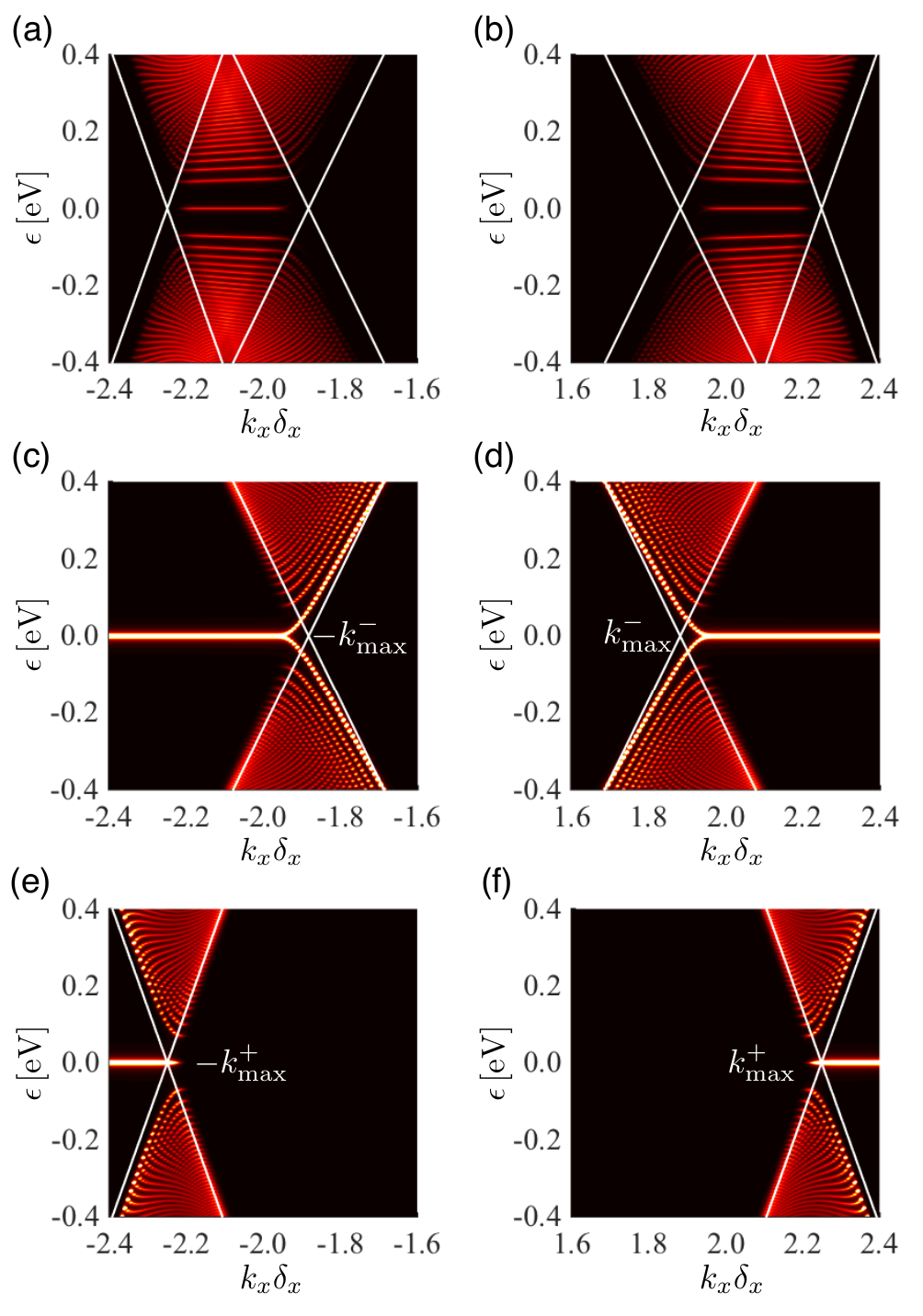}
\caption{Spectral functions of a bent graphene nanoribbon of width $W=192\,\text{nm}$ and bend curvature $\lambda=0.642\,\mu\text{m}^{-1}$. (a, c, e) and (b, d, f) plot the spectral functions in the vicinity of the left and right projected Brillouin zone corners, respectively. (a, b) The bulk spectral functions with projected Dirac cones labeled by the white curves. The states bounded between the projected Dirac cones are the strain-induced pseudo Landau levels. (c, d) The spectral functions of the stretched marginal region, whose hosted energy bands are trapped in the inner projected Dirac cones (white curves) with their Dirac points pinned at $\pm k_{\text{max}}^-$. (e, f) The spectral functions of the compressed marginal region, the energy bands of which are trapped in the outer projected Dirac cones (white curves) with their Dirac points located at $\pm k_{\text{max}}^+$.}\label{fig9}
\end{figure}

We first calculate the spectral function in the bulk of the bent graphene nanoribbon at low energies [Figs.~\ref{fig9}(a) and~\ref{fig9}(b)] and find a set of slightly dispersive bulk bands, which are the strain-induced pseudo Landau levels, bounded between the projected Dirac cones [white curves in Figs.~\ref{fig9}(a) and~\ref{fig9}(b)] characterized by $\epsilon_{\text{max}}^{\text{DC}} = \pm \hbar \tilde v_x^\eta (q_x - \eta \tfrac{g}{2a} \lambda y)|_{y=\pm W/2}$ [i.e., the weak strain limit of Eq.~(\ref{Dirac_proj})]. On a phenomenological basis, the effect of the bend on a certain Dirac point is to relocate its position along the $x$ direction in a $y$ dependent fashion. The trace constituted by the displaced Dirac points at different $y \in[-\tfrac{W}{2},\tfrac{W}{2}]$ is then a flat band spreading in the vicinity of the chosen Dirac point; and thus is the zeroth pseudo Landau level by nature. For a higher pseudo Landau level, the wave function has more nodes and consequently gets less localized in the real space. Its width in the momentum space becomes narrower.  Eventually, all the pseudo Landau levels are bounded between the aforementioned two projected Dirac cones.

\begin{figure}[t]
\includegraphics[width = 8.6cm]{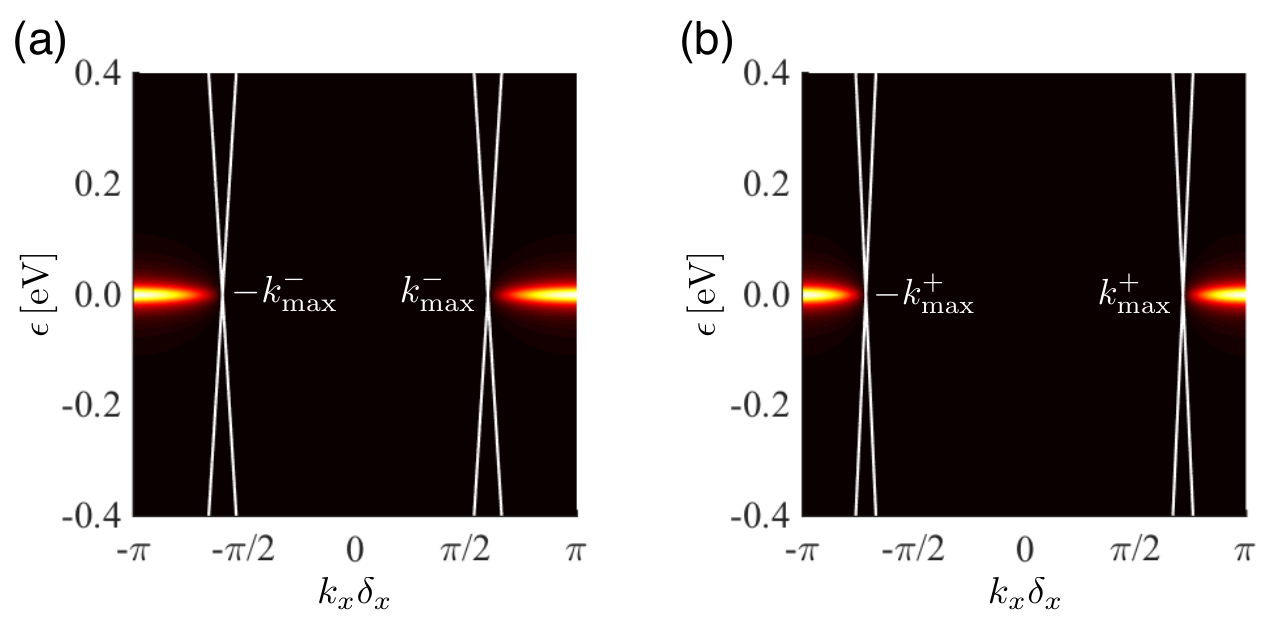}
\caption{Edge spectral functions of a bent graphene nanoribbon of width $W=192\,\text{nm}$ and bend curvature $\lambda=0.642\,\mu\text{m}^{-1}$. (a) The edge states on the stretched zigzag edge emerge from the Dirac points of the inner projected Dirac cones (white curves) at $\pm k_{\text{max}}^-$. (b) The edge states on the compressed zigzag edge emerge from the Dirac points of the outer projected Dirac cones (white curves) at $\pm k_{\text{max}}^+$.}\label{fig10}
\end{figure}

We also notice that the energy bands outside the bounded area unambiguously belong to the stretched [Figs.~\ref{fig9}(c) and~\ref{fig9}(d)] and compressed [Figs.~\ref{fig9}(e) and~\ref{fig9}(f)] marginal regions, consistent with our observation of $\bar y$ in Figs.~\ref{fig2}(a) and~\ref{fig2}(d). Most of these energy bands are strongly dispersive and spliced to the pseudo Landau levels at the boundaries of the projected Dirac cones except for a pair of longer flat bands [Figs.~\ref{fig9}(c) and \ref{fig9}(d)] degenerate with the zeroth pseudo Landau level and a pair of shorter flat bands [Fig.~\ref{fig9}(e) and \ref{fig9}(f)] connecting the two sectors of the zeroth pseudo Landau level across the Brillouin zone boundary. Further calculations of the spectral functions on the zigzag edges, i.e., the $a_1$ site and the $b_{2N}$ site in Fig.~\ref{fig1}(a), clarify that the longer (shorter) flat bands emerging from the projected Dirac points at $\pm k_{\text{max}}^-$ ($\pm k_{\text{max}}^+$) are located on the stretched (compressed) edge as illustrated in Fig.~\ref{fig10}(a) [Fig.~\ref{fig10}(b)]. We thus refer to such flat bands as the edge states, which have a topological origin (cf., Sec~\ref{sec3}), while call those dispersive bands inside the projected Dirac cones the marginal energy bands.

\bibliographystyle{apsrev4-1-etal-title_10authors}
\bibliography{bend_PR}

\begin{thebibliography}{83}
\bibitem [{not()}]{note1}%
  \BibitemOpen
  \href@noop {} {}\bibinfo {note} {According to Eq.~(\ref{pLL}), the derivative
  of $\epsilon_n(k_x)$ explicitly reads $(\frac{d\epsilon_n}{dk_x})_\mu =
  \frac{3\sqrt[4] 3 at}{8\sqrt2 g} \{ng\lambda a [e^{-2g(F_\mu-1)}-\frac{1}{4}]
  \}^{1/2} [F_\mu^2(F_\mu^2-\frac{1}{4})]^{-3/4} $, which is a decreasing
  function of the parameter $F_\mu = [4-16\mu^4/(27 t^4 n^2 g^2 \lambda^2
  a^2)]^{-1/2} $. Since $F_\mu$ decreases with an increased $\lambda$, the
  derivative $(\frac{d\epsilon_n}{dk_x})_\mu$ and consequently
  $\sigma^{xx}_b(\mu, \lambda)$ become decreasing functions of
  $1/\lambda$.}\BibitemShut {Stop}%
\end{thebibliography}
\end{document}